\documentclass[lettersize,journal]{IEEEtran}

\usepackage{cite} 
\usepackage{amsmath,amssymb,amsfonts}
\usepackage{algorithmic}
\usepackage{graphicx}
\usepackage{textcomp}
\usepackage[hyphens]{url}
\usepackage{fancyhdr}
\usepackage{hyperref}
\usepackage[final]{microtype}
\usepackage{threeparttable}

\usepackage{enumitem}
\setlist[itemize]{itemsep=1.0pt, topsep=2pt, leftmargin=*}

\usepackage{bm}
\usepackage{soul}
\usepackage{booktabs}
\usepackage[ruled,linesnumbered]{algorithm2e}

\usepackage{graphicx}
\usepackage{textcomp}
\usepackage{xspace}
\usepackage{multirow}
\usepackage{wrapfig}

\usepackage[aboveskip=1.5pt]{subcaption}

\usepackage{tikz}
\usepackage[dvipsnames]{xcolor}

\newcommand{\eg}{\textit{e}.\textit{g}.\xspace}

\newcommand*\circled[1]{\tikz[baseline=(char.base)]{
            \node[shape=circle,fill,inner sep=1.2pt] (char) {\textcolor{white}{#1}};}}

\usepackage{titlesec}
\titlespacing{\section}
{0pt}{1.0ex plus 0.6ex minus .2ex}{.4ex plus .1ex}
\titlespacing{\subsection}
{0pt}{.3ex plus .1ex minus .0ex}{.1ex plus .0ex}
\titlespacing{\subsubsection}
{0pt}{0ex plus 0ex minus .0ex}{0.ex plus .0ex}

\newcommand{\Design}{{Proxima}\xspace}

\begin{document}

\title{\Design: Near-storage Acceleration for Graph-based Approximate Nearest Neighbor Search in 3D NAND } 


\author{Weihong Xu, Junwei Chen, Po-kai Hsu, Jaeyoung Kang, Minxuan Zhou, Sumukh Pinge,\\ Shimeng Yu,~\IEEEmembership{Fellow,~IEEE}, and~Tajana Rosing,~\IEEEmembership{Fellow,~IEEE} \vspace{-0.9cm} 

\thanks{
    Weihong Xu, Junwei Chen, Minxuan Zhou, Sumukh Pinge, and Tajana Rosing are with Department of Computer Science and Engineering, University of California San Diego, La Jolla, CA 92093, USA. E-mail: \href{mailto:wexu@ucsd.edu}{wexu@ucsd.edu} 
}

\thanks{
    Po-kai Hsu and Shimeng Yu are with the School of Electrical and Computer Engineering, Georgia Institute of Technology, Atlanta, GA 30332, USA.
}

}


\IEEEtitleabstractindextext{%
\begin{abstract}
    Approximate nearest neighbor search (ANNS) plays an indispensable role in a wide variety of applications, including recommendation systems, information retrieval, and semantic search. Among the cutting-edge ANNS algorithms, graph-based approaches provide superior accuracy and scalability on massive datasets. 
    However, the best-performing graph-based ANNS solutions incur tens of hundreds of memory footprints as well as costly distance computation, thus hindering their efficient deployment at scale. 
    The 3D NAND flash is emerging as a promising device for data-intensive applications due to its high density and nonvolatility. 
    In this work, we present the near-storage processing (NSP)-based ANNS solution \Design to accelerate graph-based ANNS with algorithm-hardware co-design in 3D NAND flash. \Design significantly reduces the complexity of graph search by leveraging the distance approximation and early termination. 
    On top of the algorithmic enhancement, we implement the \Design search algorithm in 3D NAND flash using the heterogeneous integration technique. To maximize 3D NAND's bandwidth utilization, we present a customized dataflow and optimized data allocation scheme. 
    Our evaluation results show that, compared to graph ANNS on CPU and GPU, \Design achieves a magnitude improvement in throughput or energy efficiency. \Design yields $7\times$ to $13\times$ speedup over existing ASIC designs. 
    Furthermore, \Design achieves a good balance between accuracy, efficiency, and storage density compared to previous NSP-based accelerators. 
\end{abstract}

}

\maketitle
\IEEEdisplaynontitleabstractindextext

%
\IEEEpeerreviewmaketitle


\section{Introduction}
\label{sec:introduction}

\IEEEPARstart{N}{earest} neighbor search (NNS) is a fundamental workload that plays an important role in a wide variety of applications, such as recommendation systems and media data retrieval. 
The state-of-the-art NNS system adopts semantic-based search for unstructured data such as images, texts, videos, and speech.
The feature vectors of product catalogs are first generated using {neural embedding techniques} that can effectively capture the semantics of objects. 
Then the recommendation results are returned by finding products whose embeddings are closest to the embedded search query. 
For example, Amazon~\cite{linden2003amazon} builds semantic search engines to recommend products.

While exhaustive search is the most accurate way to perform NNS, the ever-expanding data volume makes it impractical for meeting low-latency requirements at scale. This is because it requires an expensive distance computation in high-dimensional space and linear search time. To address this issue, modern NNS systems~\cite{nsg, faiss} adopt approximate NNS (ANNS) schemes, which provide significantly lower query latency by approximating NNS results. ANNS achieves high efficiency mainly by reducing search space, distance computation, and data access. This relied on the pre-built indices that heuristically guide the search process. 
Popular indexing methods include hashing~\cite{he2013kmean_hash}, inverted file (IVF)~\cite{ivf}, and vector compression~\cite{product_quantization}. 
State-of-the-art ANNS tools with advanced indexing, such as Facebook's Faiss~\cite{faiss}, and Microsoft's DiskANN~\cite{diskann}, can deliver millisecond query latency even on large-scale datasets. 
Meanwhile, several hardware designs~\cite{anna} are presented to further push efficiency and performance beyond CPU and GPU.

The aforementioned designs improve both computational and memory efficiency. But they still suffer from significant accuracy degradation due to lossy compression and approximation. The experiments on FAISS~\cite{faiss_index_1G} show that even the carefully optimized IVF-PQ only achieves approximately 80\% recall on datasets with 10 million items. 
In comparison, recent graph-based ANNS algorithms~\cite{hnsw,nsg,diskann} demonstrate superior performance and complexity trade-offs.
HNSW~\cite{hnsw} and DiskANN~\cite{diskann} achieve a recall rate of $>98$\% with promising throughput across various large datasets.
As such, graph-based methods have been integrated into state-of-the-art ANNS tools~\cite{faiss,diskann,hnsw,nsg}.

However, implementing graph-based ANNS efficiently poses great challenges due to two reasons:
First, graph-based ANNS is notorious for massive memory footprint because both the raw data and the graph index need to be stored for the best accuracy. Advanced graph ANNS tools, such as HNSW~\cite{hnsw}, HM-ANN~\cite{hm_ann}, and NSG~\cite{nsg}, require 300-700GB memory to store the data structure for billion-scale datasets. Hence, it is challenging to store and process the large-scale graph index on CPU or GPU memory. 
Second, the query search over the graph is, in essence, a best-first (BF) graph traversal characterized by random data access. This nature entails low parallelism and expensive distance computation. 
Our profiling results in Section \ref{subsec:motivation} show that the irregular data access during graph traversal incurs 80\% to 95\% cache miss rate. Additionally, the graph search requires thousands of distance computations for high-dimensional data. Data fetching and distance computation consume up to 80\% of the total runtime.

Several attempts have been made to efficiently boost graph-based ANNS. NSG~\cite{nsg} and GGNN~\cite{ggnn} distribute the large graph index to multiple machines~\cite{nsg} or GPUs~\cite{ggnn}. However, this approach dramatically increases the deployment cost while sacrificing energy efficiency due to the communication overhead. 
SONG~\cite{SONG} optimizes the graph data organization for GPU to exploit GPU's computation parallelism. They use hashing to fit large indexes into limited GPU memory, but significant performance degradation occurs in a high-recall regime due to the hash approximation. 
Besides, several works have offloaded the memory-intensive graph index from DRAM to slower but cheaper and denser memory devices. 
Emerging non-volatile memories, like solid state drives (SSDs) and Optane~\cite{hm_ann}, are the ideal devices to store the graph index. 
DiskANN~\cite{diskann} offloads the graph index from DRAM to SSD whereas the host memory only caches frequently-accessed data. 
HM-ANN~\cite{hm_ann} builds a heterogeneous memory architecture that combines DRAM and non-volatile Optane memory. 
However, the main drawback of these solutions is that the data in storage needs to travel through multiple-level memory hierarchies for computing, thus incurring costly data movement.

The near-data processing paradigm offers a compelling opportunity to accelerate graph-based ANNS by moving computation closer to where data resides, thereby increasing available bandwidth and enabling a more energy-efficient datapath. Existing efforts can be broadly categorized into near-DRAM and near-storage designs. \textbf{Near-DRAM accelerators} such as Pyramid~\cite{zhu2023processing} improve internal memory bandwidth, but they still store full-precision vectors and large-scale graphs in DRAM. As a result, their scalability remains constrained by DRAM capacity and cost, especially for billion-scale graph indices with massive memory footprints.
\textbf{Near-storage processing (NSP) designs}, such as NDSearch~\cite{wang2024ndsearch}, VStore~\cite{vstore}, and REIS~\cite{chen2025reis}, leverage the ultra-high density of 3D NAND flash~\cite{3DNAND-CIM} to reduce DRAM storage cost. However, our profiling analysis (see Section~\ref{subsec:motivation}) shows that these architectures continue to face fundamental SSD bottlenecks, including channel-level and controller-level latencies. Because computation occurs above the flash array, they still rely on coarse-grained page transfers, shared channel buses, and limited in-SSD DRAM, ultimately restricting their achievable parallelism and latency.

To fill this gap, we present an algorithm–hardware co-design, \Design, to address the key NSP challenges in accelerating graph-based ANNS. \Design is a 3D-NAND-based near-storage accelerator that leverages heterogeneous integration technology~\cite{shim2021system} (hybrid bonding) to place compute units directly adjacent to the memory array, enabling fine-grained, die-level access. This architectural approach removes SSD channel bottlenecks and supports low-latency, highly parallel graph traversal directly within the flash stack.
However, deploying graph-based ANNS in 3D NAND via hybrid bonding alone is insufficient to achieve satisfactory query latency, as the intrinsic cost of random graph traversal, distance computation, and index access still dominates end-to-end performance. To overcome these limitations, we co-design both the search algorithm and hardware architecture, and propose several key optimizations. Together, these techniques significantly reduce memory traffic, improve data locality, and minimize redundant computations, enabling efficient graph search within the 3D NAND substrate.
Our contributions are summarized as follows:
\begin{itemize}[leftmargin=*]
    \item We provide an in-depth profiling and analysis for graph-based ANNS and identify its inefficiencies on CPUs and GPUs. Based on the analysis, we design the novel \Design graph search algorithm that improves throughput and efficiency by combining product quantization (PQ) and early termination.
    \item We leverage the heterogeneous integration techniques to devise \Design, a NSP accelerator in 3D NAND tailored for graph-based ANNS workloads. We present various ASIC designs to efficiently implement the proposed graph search algorithm. 
    \item We introduce various types of data mapping optimizations for \Design, which reduces query latency while providing scalable approaches to handle different dataset sizes. The experiments show that the proposed hot node repetition achieves $3\times$ latency reduction without additional hardware.
    \item \Design achieves up to $15\times$ speedup and energy efficiency improvements compared to state-of-the-art hardware solutions~\cite{anna,vstore,ggnn}.
\end{itemize}

\begin{figure}[t]
    \centering
    \includegraphics[width=0.9\linewidth]{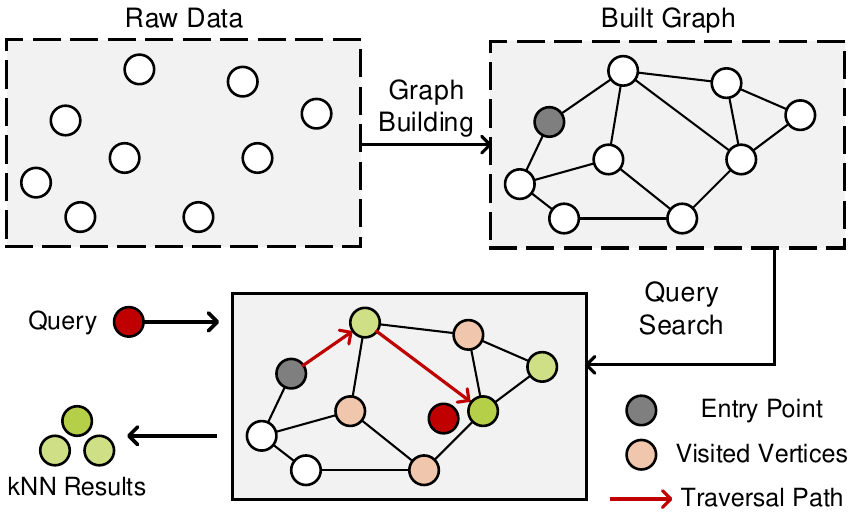}
    \caption{Illustration of graph-based ANNS.}
    \label{fig:graph_ann_example}
\end{figure}

\section{Background and Motivations}
\subsection{Approximate Nearest Neighbor Search}
\noindent
\textbf{Exact $k$-NN Search.}
The $k$-nearest neighbor ($k$-NN) search retrieves $k$ nearest neighbors, $\mathcal{R}$ from a dataset $\mathcal{X}=\{x_1,\ldots, x_N\}$, that have the smallest distances to the given query $q$. The $k$-NN search for the $D$-dimension vector in Euclidean space ${x}_i, q \in \mathbb{R}^{D}$ is given by:
\begin{equation}\label{eq:ann}
    \mathcal{R} = \underset{\mathcal{R} \subseteq \mathcal{X}, |\mathcal{R}|=k}{\arg \min}\ dist(q, x),
\end{equation}
where $dist(\cdot, \cdot)$ denotes the distance between two given vectors, \eg, Euclidean, cosine, or inner product.

\noindent
\textbf{Approximate $k$-NN Search.}
The exact $k$-NN search using exhaustive search requires $\mathcal{O}(N\cdot D)$ complexity, thus is computationally inefficient as the data size $N = |\mathcal{X}|$ becomes million or billion-scale. This makes exact $k$-NN search impractical to implement in real-time systems. 
Instead of precisely retrieving $k$ NNs, state-of-the-art tools~\cite{faiss,hnsw} relax the exact search conditions and instead retrieve the approximate $k$ NNs expressed as $\hat{\mathcal{R}}$. 
These approximate $k$-NN search algorithms effectively reduce the search complexity by only visiting a small portion of the dataset. 
The good trade-off between search complexity and accuracy heavily depends on the pre-built data index that guides the search process.  Popular data indexing approaches adopted by existing large-scale NN search libraries include IVF~\cite{ivf}, graph~\cite{diskann,nsg,hnsw}, and hashing~\cite{he2013kmean_hash}.


\noindent
\textbf{Evaluation Metrics.}
Recall, query latency, and throughput are the three key metrics to evaluate the ANNS performance. The recall measures the overlap between the approximate $k$-NN set $\hat{\mathcal{R}}$ and the exact $k$-NN set is $\mathcal{R}$, which is computed by:
\begin{equation}\label{eq:ann_recall_qps}
    Recall(\hat{\mathcal{R}}, \mathcal{R}) = \dfrac{| \hat{\mathcal{R}} \cap \mathcal{R}|}{k}.
\end{equation}

Query latency measures the response latency of a given ANNS system, while throughput is measured in terms of query per second (QPS). Therefore, the design goal is to obtain high QPS and low latency while providing satisfactory recall.

\subsection{Graph-based ANNS}
The experiments~\cite{faiss_index_1G} using Facebook's FAISS~\cite{faiss} show IVF~\cite{ivf} and hashing~\cite{he2013kmean_hash} methods are memory-efficient but the yield recall saturates around 80\% on 10M and 100M-scale datasets. 
In comparison, graph-based methods~\cite{hnsw,diskann,nsg} demonstrate superior performance with polylogarithmic search and graph building complexity. The graph-based ANNS includes two phases: 1. graph building and 2. search. The graph building generates a sparse \textit{proximity graph}, $G(V, E)$, as the data index. 
Each data point $x_i\forall \mathcal{X}$ is uniquely represented by vertex $v_i \in V$ over the graph. The edge $e \in E$ represents the neighborhood relationships for the connecting vertices.

Although existing graph-based ANNS algorithms~\cite{hnsw,diskann,nsg} impose diverged constraints during graph building, most of them share a similar heuristic searching procedure. The search flow is a best-first traversal illustrated in Fig.~\ref{fig:graph_ann_example}. 
As a new query $q$ comes, the search process starts from a pre-defined \textit{Entry Point} ($v_s$) and greedily traverses the graph to reach the nearest neighbors of $q$. A candidate list $\mathcal{L}$ is maintained to store (distance, id) pairs of the best evaluated vertices, which are sorted in ascending order of their distances to $q$. $\mathcal{L}$ has a predefined size $L$ and is intialized with (dist($v_s$,$q$), $v_s$). The search process iteratively "evaluates" the first unevaluated candidate in $\mathcal{L}$ by visiting its neighborhood and computing the distances between its neighbors and $q$. These neighbors are inserted into $\mathcal{L}$ along with their distances. Then $\mathcal{L}$ is sorted and only keeps the top $L$ nearest candidate to $q$. This search process is repeated until all candidates in $\mathcal{L}$ have been evaluated. Then the first $k$ candidates in $\mathcal{L}$ are returned as an approximation of $k$ nearest neighbors to $q$.  
The candidate list size $L$ can be used to control the search accuracy. In particular, with a higher $L$, more vertices in the graph are evaluated before the search terminates. Therefore, the search returns more accurate results. 


\subsection{3D NAND-based Near-Storage Processing}

\begin{figure}[t]
    \centering
    \includegraphics[width=0.75\linewidth]{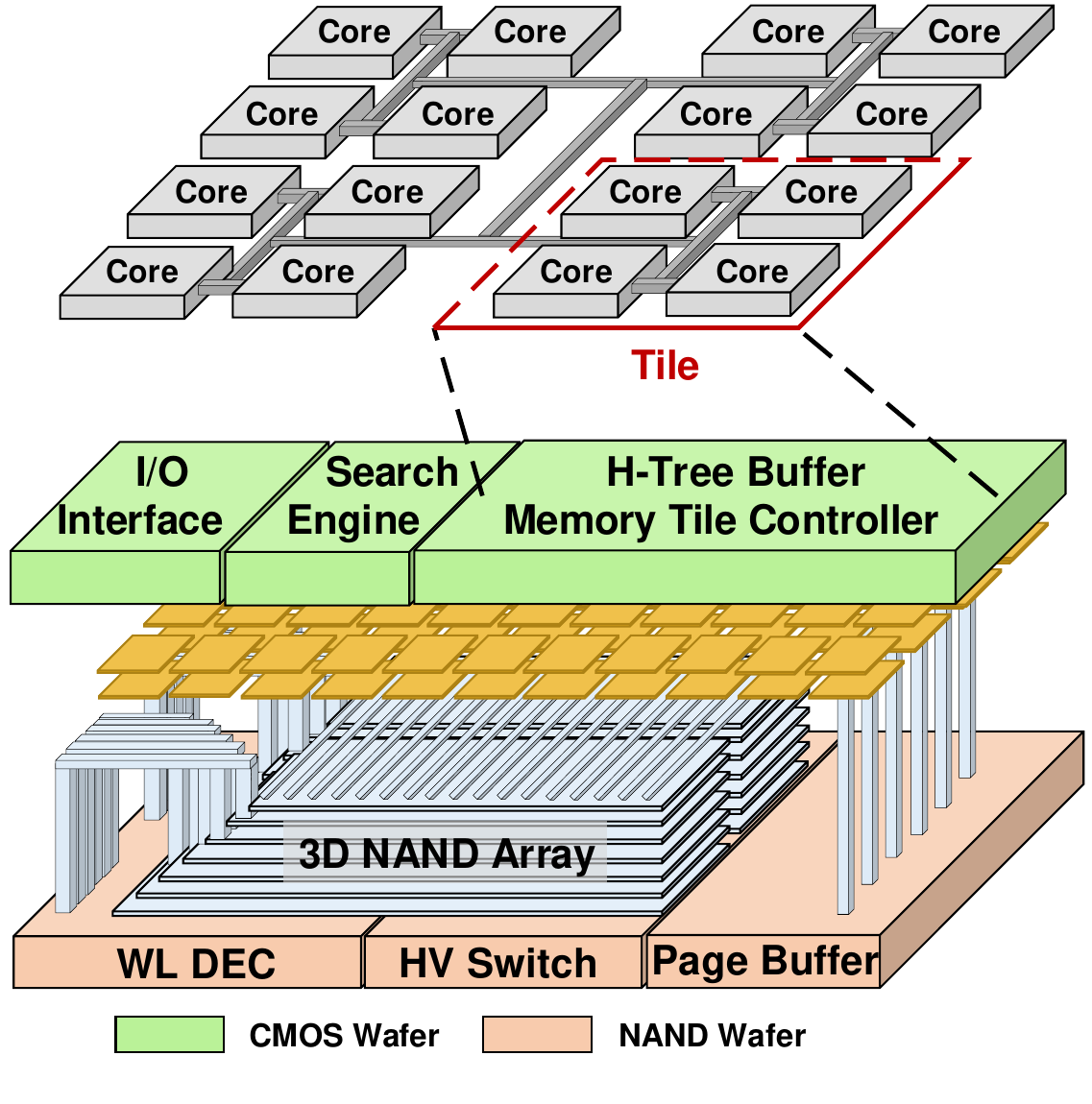}
    \caption{\Design's near storage architecture with 3D NAND and heterogeneous integration techniques.}
    \label{fig:3d_nand_arch}
\end{figure}

\noindent
\textbf{Near-storage Processing (NSP).}
As emerging algorithms become increasingly IO-bounded, the growing demand for large-scale data processing presents significant challenges for conventional von Neumann architectures, which require massive data transfers between the processor and off-chip memory. Limited bandwidth and expensive data movement hinder the computing system scaling for data-intensive workloads, causing the well-known "memory wall" problem. To overcome this limitation, in/near-memory processing architectures based on emerging non-volatile memory (NVMs)~\cite{PRIME} and volatile dynamic random access memory (DRAM)~\cite{Ambit} 
have shown promise in building optimized accelerators for various big data applications. However, these devices may not be suitable for larger planar array setups due to their low on/off current ratio, high on-current, and relatively lower density to support large datasets. Recently, 3D NAND flash memory has emerged as a promising in-storage processing (ISP) candidate~\cite{3D-FPIM,3DNAND-CIM,shim2021system}, offering high density and bandwidth. However, ISP designs using 3D NAND flash memory suffer from: 1. device and circuit non-idealities that reduce computing accuracy, 2. variations in process, voltage and temperature variations, and 3. device aging~\cite{3DNAND-aging}. To avoid this issue, near-storage processing (NSP) as an alternative could still provide an energy-efficient and low-latency solution to accelerate large-scale data processing without offloading massive data from the 3D NAND chip.

\noindent
\textbf{Heterogeneous Integration for NSP.}
Although NSP avoids massive data transfers, the integration of processing elements could be area costly. Heterogeneous integration~\cite{shim2021system} using CMOS under Array (CUA)~\cite{CUA} and Cu-Cu bonding~\cite{YMTC} can address the cost and area concerns associated with integrating processing elements onto a 3D NAND chip. As shown in Fig.~\ref{fig:3d_nand_arch}, CUA overlaps memory peripherals under the array, reducing the area of a single tier. 
The CMOS wafer and NAND wafer can be manufactured independently using different technology nodes~\cite{YMTC,shim2021system}. After finishing the manufacturing of CMOS and NAND wafers, the high-density inter-chip Cu-Cu bonding connects the processing elements on the CMOS wafer to the 3D NAND wafer, providing seamless integration. As a result, NSP with heterogeneous integration can provide a compact solution for large-scale data processing with improved performance. This approach opens new opportunities for the development of low-power, high-performance, and compact data processing systems for various applications.

\subsection{Motivations}\label{subsec:motivation}

\begin{figure}[t]
    \centering
    \includegraphics[width=\linewidth]{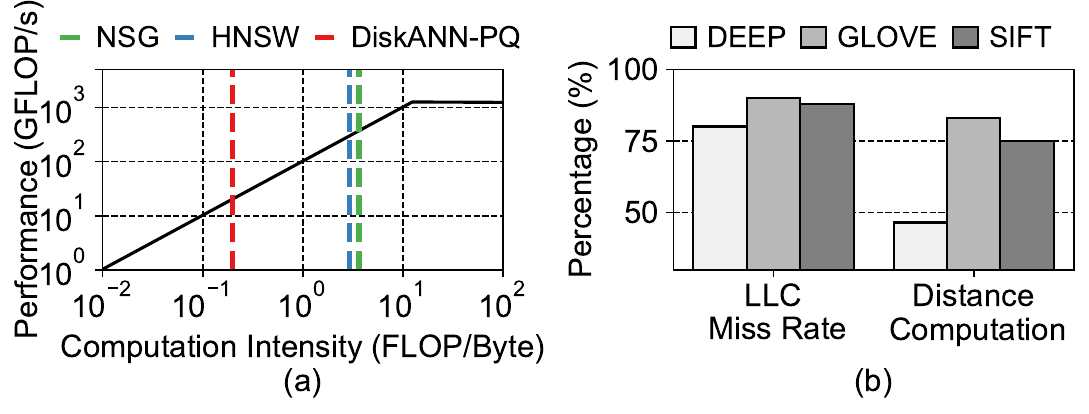}
    \caption{Profiling results for graph ANNS tools: (a) Roofline model on AMD EPYC 7543 CPU. (b) LLC miss rate and distance computation overhead for HNSW~\cite{hnsw}.}
    \label{fig:roofline_profiling}
\end{figure}

We profile the state-of-the-art graph ANNS tools (NSG~\cite{nsg}, HNSW~\cite{hnsw}, and DiskANN~\cite{diskann}) on a 16-core CPU with 64GB memory. We choose three popular datasets, SIFT~\cite{sift_1B}, GLOVE~\cite{glove}, and DEEP~\cite{deep_1B}. We identify the graph-based ANNS faces the following challenges:
\begin{itemize}[leftmargin=*]
    \item \textbf{Challenges 1: Costly Memory Footprint.} 
    Graph-based ANNS must store both the graph index and the raw data vectors. The adjacency-list index requires $\mathcal{O}(b_{index}\cdot N\cdot R)$ memory, where $N$ is the number of vertices, $R$ the maximum degree, and $b_{index}$ the bit width per index. Raw vectors require $\mathcal{O}(b_{raw}\cdot N\cdot D)$, with $D$ denoting dimensionality.
    In practice, the graph index often exceeds the raw data size because $b_{index}$ is typically 32 bits and $R$ ranges from 32 to 96. For billion-scale datasets, the index alone can exceed 203 GB and is larger than the 178 GB needed for raw data, making it difficult to store and process the full graph in DRAM or GPU memory. This motivates the need for the alternative high-density 3D NAND memory based on NSP techniques.
    \item 
    \textbf{Challenges 2: Random Access and Long Latency.} 
    Graph-based ANNS performs best-first traversal with frequent fetches of $D$-dimensional vectors and $R$ neighbor indices. As shown in Fig.\ref{fig:roofline_profiling}, these algorithms\cite{nsg,hnsw,diskann} are memory-bound and exhibit highly irregular access patterns, resulting in LLC miss rates of 80–90\% and making random access a dominant latency bottleneck.
    Existing NSP designs near SSD channels or controllers~\cite{wang2024ndsearch,vstore,chen2025reis} remain constrained by channel arbitration, controller overheads, and coarse page reads, leading to $\mu$s-level latency per traversal step. Executing ANNS directly on 3D NAND also suffers from poor locality and costly scattered accesses, which are misaligned with the fine-grained access needs of graph traversal.
    Thus, hardware–software co-design is essential. \Design adopts hybrid bonding~\cite{YMTC} and performs design-space exploration of the 3D NAND module to enable fine-grained, die-level access. Meanwhile, we introduce several data-mapping techniques (hotness-based reordering, node repetition, and gap encoding) to improve locality and reduce access overhead. Together, these co-optimized mechanisms significantly lower query latency and enable efficient graph traversal inside 3D NAND.   
    \item 
    \textbf{Challenges 3: Expensive Distance Computation.}
    Achieving high recall in graph-based ANNS requires frequent accurate distance evaluations over high-dimensional vectors, accounting for more than 50\% of runtime on CPU (Fig.~\ref{fig:roofline_profiling}-(b)). Yet many of these computations are redundant and do not affect the final results, making them a major source of inefficiency. 
    To mitigate this overhead, \Design adopts memory-efficient search techniques that reduce unnecessary distance evaluations. By combining PQ-based distance approximation with dynamic candidate lists and early termination, \Design focuses computation only on vertices likely to impact the search outcome, significantly improving efficiency while preserving recall.
\end{itemize}


\begin{figure}[t]
    \centering
    \includegraphics[width=0.95\linewidth]{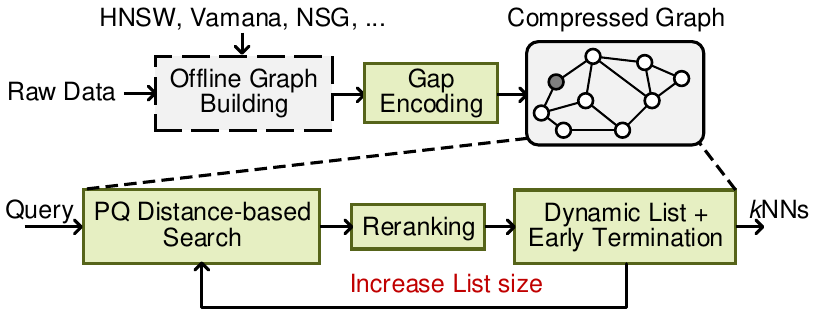}
    \caption{\Design graph search algorithms.}
    \label{fig:alg_flow}
\end{figure}

\section{\Design Graph Search}\label{sec:algorithm}
According to Section \ref{subsec:motivation}, ANNS graph search is dominated by costly distance computation and memory footprint. We present the optimized \Design graph search scheme to address these bottlenecks at the algorithm level.

\subsection{Overview}

Fig.~\ref{fig:alg_flow} shows the overall flow of \Design search optimization. 
\Design graph search is general and can be applied to graphs generated by various graph building algorithms, such as HNSW~\cite{hnsw}, DiskANN~\cite{diskann}, and NSG~\cite{nsg}. 
Before query search, the gap encoding in \Design first compresses the built graphs to reduce the index size. 
During query search, we propose three strategies, including product quantization (PQ) distance-based search, accurate distance-based reranking, and dynamic list with early termination, to perform accurate and low-complexity search.

\begin{figure}[t]
    \centering
    \includegraphics[width=\linewidth]{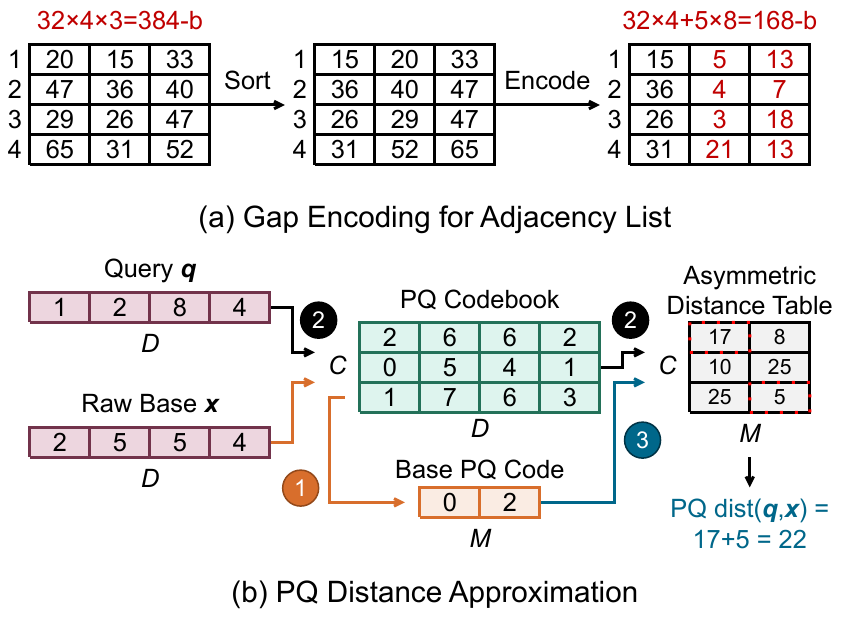}
    \caption{(a) Gap encoding and (b) product quantization (PQ) distance approximation in \Design.}
    \label{fig:gap_pq}
\end{figure}

\SetKwComment{Comment}{/* }{ */}
\begin{algorithm}[t]
    \small
    \caption{\Design Search Algorithm}\label{alg:ann_search}
    \KwData{Graph $G(V,E)$; Query $q$; Entry point $v_{s}$; Repetition rate $r$; Search step $T_{step}$; Candidate list size $L$; PQ threshold $beta$.}
    \KwResult{$k$-NN results for query $q$.}
    Candidate set $\mathcal{L} = \{(\hat{dist}_{s}, v_{s})\}$\;
    Evaluated set $\mathcal{E} = \{ \}$\;
    \While{$T\le L$}{
        $(\hat{dist}', v')$ = the first candidate in $\mathcal{L}$ with $v' \notin \mathcal{E}$\;
        $\mathcal{E} = \mathcal{E} \cup \{ v' \}$\;
        \ForEach{$n\in$ neighbors of $v'$}{
            $\hat{dist}_{n} = \hat{dist}(q, n)$ \Comment*[r]{PQ Distance}
            $\mathcal{L} = \mathcal{L} \cup \{ (\hat{dist}_{n}, n) \}$\;
        }
        Sort $\mathcal{L}$ based on $\hat{dist}$ and keep top $L$ candidates\;
        \If{$\forall a \in$ top $T$ candidates of $\mathcal{L}, a \in \mathcal{E}$}{
            Rerank top $T$ candidates in $\mathcal{L}$ \;
            \If{early termination check($r$) == True}{
                 break \Comment*[r]{Early Termination}
            }
            $T = T + T_{step}$ \Comment*[r]{Dynamic List}
        }
    }
    \ForEach{$c\in$ $\mathcal{L}$ and $\hat{dist}_{c}<\hat{dist}_{\mathcal{L}[T]}*beta$}{
        Rerank $c$ \Comment*[r]{Reranking}
    }
    \Return Top-$k$ reranked candidates
\end{algorithm}

\subsection{PQ Distance-based Search}
Graph ANNS algorithms \cite{hnsw, nsg, diskann} incur hundreds to thousands of distance computations, but most do not contribute to the final results. 
A natural idea is to approximate them. \Design traverses the graph using the approximate distance obtained by PQ~\cite{product_quantization}. 
PQ splits vectors into $M$ subdimensions and encodes each vector $x$ into a $M\cdot \log_{2}C$-bit code, where $C$ is the number of centroids of each subdimension from $k$-means. Each subvector of $x$ is quantized into its $k$-means centroid and encoded with centroid index of $\log_{2}C$ bits.
PQ approximates the distance $\hat{dist}$ between query $q$ and $x$:
\begin{equation}\label{eq:pq_dist}
    \hat{dist} \left( q, x \right) = \sum_{i=1}^{M} \mathbf{ADT}_i[{c}_i],
\end{equation}
where $\mathbf{ADT}_i$ is the asymmetric distance table (ADT) computed as we receive $q$, which holds distances between subvector $q_i$ of $q$ and the $C$ centroids of subdimension $i$; ${c}_i$ is the centroid index of subvector $x_i$ of $x$. PQ approximate distance $\hat{dist}$ is computed by summing the distance between each subvector of $q$ and the quantized subvector of $x$. Therefore, a total of $M$ LUTs (Look Up Table) and additions are needed for each distance computation. Fig.~\ref{fig:gap_pq}-(b) depicts an example of computing the ADT and final PQ distance for $D=4, M=2, C=3$. Note that \Design only uses PQ during the search process. The graph index is built using existing algorithms \cite{hnsw, nsg, diskann} with full-precision coordinates to ensure correct structure.

\subsection{Accurate Distance-based Reranking}\label{subsec:rerank}
When the search ends, we obtain a list $\mathcal{L}$ of best-matched vertices found during graph traversal using PQ, but we still need to rerank these vertices to return the actual top $k$ results in $\mathcal{L}$. Compared to the thousands of accurate distance computations in traditional graph searching algorithms, the cost of reranking (typically around one hundred) is trivial. 

Although PQ with reranking has been a common technique for graph-based ANNS~\cite{diskann, faiss}, they have not taken full consideration of the inaccuracy of PQ. Let $\hat{dist}_{\mathcal{L}[T]}$ be the distance of the most distant candidate in $\mathcal{L}$, then vertices with an accurate distance less than $\hat{dist}_{\mathcal{L}[T]}$ but a PQ distance greater than $\hat{dist}_{\mathcal{L}[T]}$ are discarded merely because of PQ's inaccuracy. To address this issue, we propose an optimized reranking strategy by introducing a new parameter $beta$ (PQ error ratio) into Algorithm~\ref{alg:ann_search} (line 19) and nesting the original search list of size $T$ inside a larger list of size $L$. $beta$ can be chosen through empirical evaluation. For example, by sampling base vertices as queries and constructing 32 bytes PQ code, we observe that 99 percent of PQ distances for the SIFT dataset are within 1.06 ($beta$) times of their accurate distances. Then, we rerank all vertices with PQ distances less than $\hat{dist}_{\mathcal{L}[T]}*beta$. This is possible because we keep more vertices in the larger list of size $L$. We may need a few accurate distance computations, but we can improve the recall by up to 10\% over DiskANN\cite{diskann} at a low recall with negligible impact on QPS. (Fig.~\ref{fig:software_recall_qps}).

\subsection{Dynamic List and Early Termination}\label{subsec:dynamic_et}
\begin{figure}[t]
    \centering
    \includegraphics[width=\linewidth]{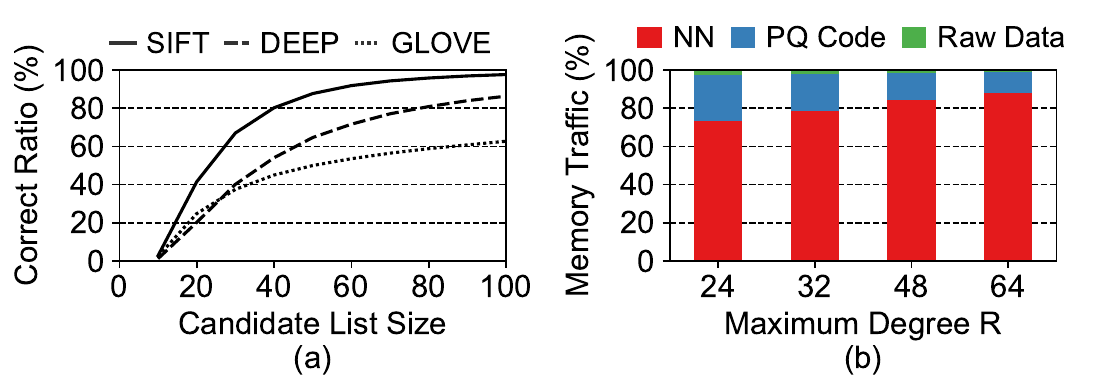}
    \caption{(a) Search convergence trend for GLOVE, DEEP10M, SIFT. (b)Memory traffic breakdown using different degrees $R$.}
    \label{fig:convergence_traffic}
\end{figure}

We propose a Dynamic List and Early Termination strategy to make use of the information during graph traversal, as shown in Algorithm~\ref{alg:ann_search}. We observe that most queries converge (find their true k-NNs) at a small $T$ (candidate list size). Increasing $T$ further will not improve the recall of these queries, but only increase the computation cost. Fig.~\ref{fig:convergence_traffic}-(a) shows this trend for DiskANN~\cite{diskann}. We see a rapid increase of convergence ratio at small $T$'s. For datasets like GLOVE having low recall even at big $T$'s, we will expect less queries to converge at a given $T$, but the general trend is the same. Based on this, we design a novel early termination technique that uses a dynamic list to iteratively adjust $T$ (line 16) during graph traversal. In each iteration, we compare the top-$k$ new and old reranked candidates, and terminate the search if they are the same for $r$ consecutive iterations (line 12-14); otherwise, we increase $T$ by $T_{step}$ (line 16) and continue. We control the early termination outcome by setting $r$ and $T_{step}$. We store the computed distances to amortize the overhead of accurate distance computations in each iteration. We also keep $\mathcal{L}$ at a relatively large size $L$ to preserve the useful information at small $T$'s. This also allows us to apply the optimized reranking scheme in Section \ref{subsec:rerank}. We combine early termination with optimized reranking and found around 10\% reduction in distance computations at the same recall for all 6 datasets in Table.\ref{table:dataset}, compared to optimized reranking alone. This suggests that early termination can scale well to datasets of different sizes and distributions.

\section{\Design Accelerator}\label{sec:arch}

Although \Design search optimization effectively improves graph search, it is still limited by costly data movement between SSD, host memory, and cache. 
For better throughput and efficiency, we architect the near-storage accelerator using 3D NAND and heterogeneous integration in this section to implement the \Design graph search algorithm. 

\begin{figure}[t]
    \centering
    \includegraphics[width=0.9\linewidth]{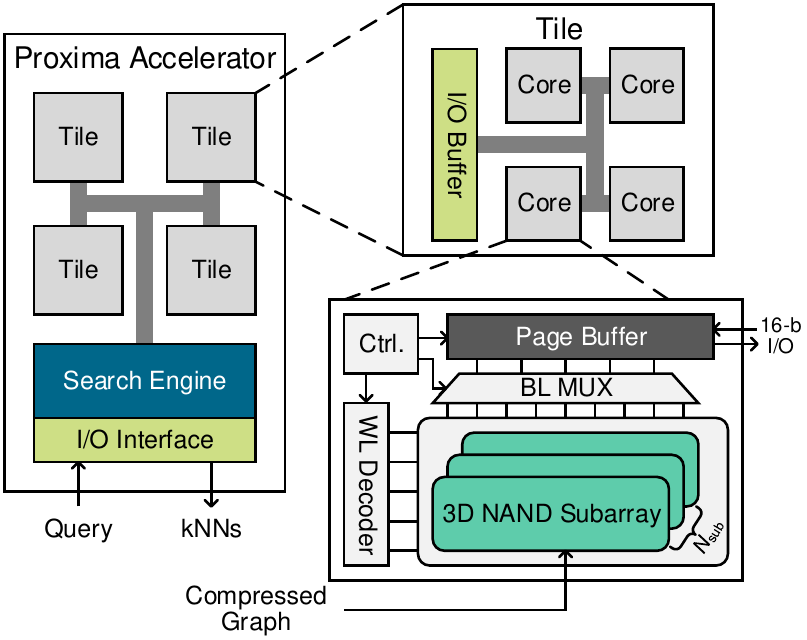}
    \caption{\Design accelerator in 3D NAND flash.}
    \label{fig:arch}
\end{figure}

\subsection{Architecture Overview}

Fig.~\ref{fig:arch} illustrates the architecture of \Design accelerator consisting of two main parts: \textit{3D NAND Tiles} and \textit{Search Engine}. \Design is a standalone near-storage accelerator (similar to~\cite{vstore,ice}) that can realize both data storage and efficient ANNS function. 
This is achieved by utilizing the advanced heterogeneous integration technique~\cite{YMTC}, which connects the 3D NAND wafer and CMOS wafer for better efficiency and storage capacity. 

\noindent
\textbf{3D NAND Tiles} are implemented on the 3D NAND wafer and are used to store three types of graph data: 1. raw data, 2. raw data's PQ codes, and 3. compressed graph NN indices. To ensure the architectural extensibility, the two-level spatial hierarchy in previous works~\cite{3D-FPIM,shim2021system} is adopted to construct 3D NAND tiles, where there are total of $N_{tiles}$ tiles and each tile includes $N_{core}$ cores. The tiles and cores at the same hierarchy are connected via the tile or core H-tree bus, respectively. Inside each core, there are $N_{sub}$ blocks of 3D NAND subarrays with peripheral circuits. 
To boost data access during graph search, we customize the 3D NAND core by adding a BL MUX between BLs and page buffer. Meanwhile, a smaller array size is chosen for lower read latency. The details are presented in Section~\ref{subsec:3d_nand_flash}.

\noindent
\textbf{Search Engine} is manufactured on the CMOS wafer to maximize the logic density. Then it is connected to the tile controller via \textit{H-tree bus} and further connected to the memory array through Cu-Cu bonding~\cite{YMTC}. 
The search engine efficiently implements \Design graph search algorithm and is used for: 1. scheduling and processing incoming queries, 2. fetching graph data from the 3D NAND cores. The 3D NAND flash core is the minimal unit that can be accessed by the search engine for data fetching. The design of the search engine is given in Section~\ref{subsec:search_engine}.

\subsection{Execution Flow} 
The execution of \Design accelerator is mainly composed of two steps: 1. graph data preloading and 2. query graph search. Before graph search, the raw data, raw data's PQ codes, and NN indices need to be prestored into the corresponding physical addresses using the proposed data mapping scheme shown in Section~\ref{sec:data_mapping}. 

\begin{figure}[t]
    \centering
    \includegraphics[width=0.9\linewidth]{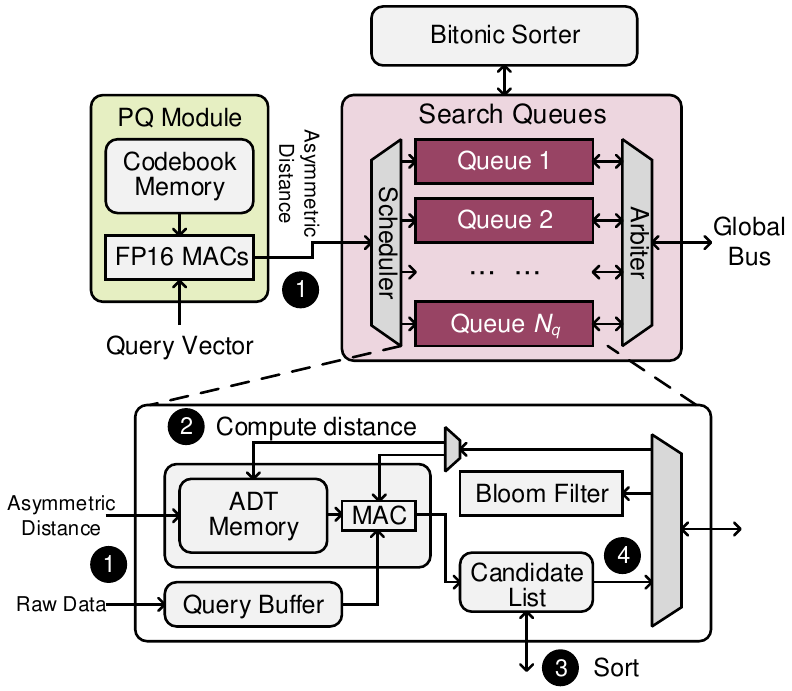}
    \caption{Internal architecture of search engine.}
    \label{fig:search_engine}
\end{figure}

After the required data have been stored, Fig.~\ref{fig:search_engine} illustrates the data flow inside the search engine during graph search. 
The entire graph search consists of four steps:
Step \circled{1} is the initial phase for new queries, where the PQ module computes the asymmetric distance table (ADT) based on the query data. The computed ADT is transmitted to the ADT memory in the target queue specified by the scheduler. In Step \circled{2}, the candidate list in the queue pops the un-evaluated vertex, and the arbiter generates the corresponding address to fetch the NN indices from 3D NAND cores (Line 4-6 in Algorithm~\ref{alg:ann_search}). The fetched NN indices first pass through the Bloom filter to update newly visited vertices and filter out those previously computed vertices. The PQ codes of those unvisited vertices are then fetched and computed by the distance computation module (Line 6-8 in Algorithm~\ref{alg:ann_search}). 

In Step \circled{3}, a sorting is needed to select the top $L$ candidates (Line 10 in Algorithm~\ref{alg:ann_search}) after all the neighbors have been visited. The implemented Bitonic sorter is a shared sorter by all search queues since it can provide sufficient throughput. After the PQ distance-based search is finished, the top vertices in the candidate list memory will be reranked using their raw data (Line 12 in Algorithm \ref{alg:ann_search}). This is illustrated by Step \circled{4} in Fig.~\ref{fig:search_engine}. Meanwhile, the candidate list also checks whether the early termination condition is satisfied.

\subsection{3D NAND Core Design}\label{subsec:3d_nand_flash}

The generic 3D NAND flash chips~\cite{kang201913,higuchi202130} used for commercial SSD are not suitable for graph ANNS. First, 3D NAND chips in SSD are optimized for storage density, thus they need large 3D NAND array sizes. As a result, the long precharging and discharging time lead to long page read latency, ranging from 15$\mu$s to 90$\mu$s~\cite{kang201913,higuchi202130}. This is unfavorable for graph traversal which needs lots of random data access and low processing latency. 
The second drawback is that the large page size (8KB to 16KB) used by existing 3D NAND flashes is too coarse for graph traversal workloads because the data reading operations of graph search (fetching raw data or graph indices) only need $<$0.5KB from the storage each time. Hence, a finer data access granularity with lower latency is needed. 
Thirdly, for attaining higher storage density, most SSDs store multiple bits per cell (MLC). However, MLC incurs one to two orders of magnitudes error rates higher than single-level cell (SLC)~\cite{RBER_SLC,RBER_MLC}. 
Our experiments in Section~\ref{subsec:analysis_error} provide numerical data to demonstrate that MLC without error correction (ECC) dramatically degrades the ANNS accuracy. 
While commercial SSDs are generally equipped with ECC modules to correct the potential bit errors when reading data~\cite{shao2017dispersed}, adding ECC to \Design is infeasible due to the large area overhead of ECC~\cite{shao2017dispersed} which may the deteriorate system efficiency and the ECC module to match the high data rate over Cu-Cu bonding between the CMOS and 3D NAND wafer.

\begin{figure}[t]
    \centering
    \includegraphics[width=\linewidth]{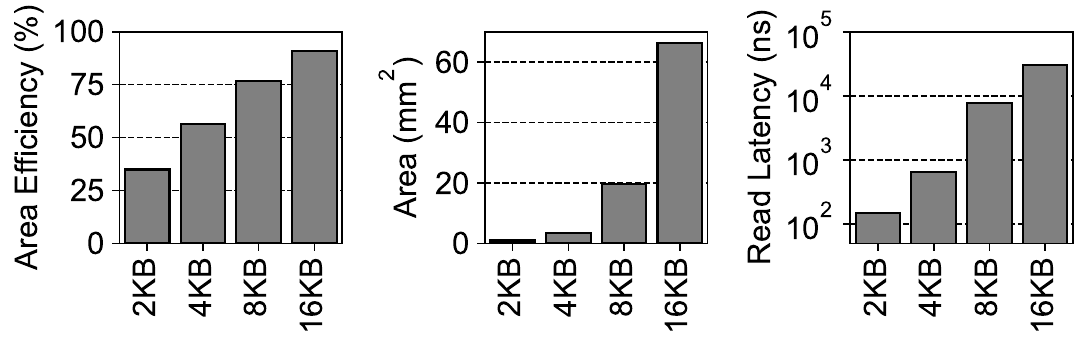}
    \caption{The density, area, and read latency trade-offs for 96-layer 3D NAND flash.}
    \label{fig:nand_tradeoff}
\end{figure}

We develop a simulator based on the 3D NAND simulator~\cite{3D-FPIM} to project the tradeoff between density, area, and read latency for 96-layer 3D NAND in Fig.~\ref{fig:nand_tradeoff}, working as the design guidance for \Design cores. 
The large page size could incur over 10$^4$ns read latency. The previous work~\cite{park2021reducing} shows the precharge and discharge processes take $\approx 90\%$ of the page read latency. The long precharge and discharge time results from the large capacitance load created by hundreds of blocks and extensive page size.  
We customize the \Design 3D NAND core to reduce the long read latency and increase data granularity. 
First, we reduce the number of bit-lines (BLs) $N_{BL}$ as well as the number of blocks $N_{sub}$ to decrease the capacitance load. Meanwhile, we also choose the proper number of select string line (SSL). Specifically, we use $N_{BL}=32768$, $N_{SSL}=4$, and $N_{block}=64$ to build each core. 
Dramatically decreasing page size also decreases the area efficiency because a large portion of NAND array is occupied by the peripheral circuits. 
To reduce the overhead of the peripheral page buffer, a BL MUX is implemented between the page buffer and 3D NAND blocks for reduced page buffer size and better data granularity. For each time, \Design's core only precharges a small portion of the BLs instead of the entire page. As a result, we use a 32:1 MUX to achieve 128B data granularity while \Design core achieves read latency $<$ 300ns. The other benefit of partial precharging is that it reduces the area overhead of the peripheral circuits in the page buffer by a factor of 32, thereby increasing the storage density.

\subsection{Search Engine Design}\label{subsec:search_engine}


\noindent
\textbf{Scheduler and Arbiter.} 
The scheduler adopts the simple Round-Robin strategy with the first-come-first-serve policy to allocate the new query to the ideal queue. The scheduler keeps track of all queues' status in an $N_q$-b buffer. 
The arbiter is used to allocate the data fetching request from queues to the target 3D NAND core. To this end, the arbiter first translates the vertex information from the queue into the associated physical address. If the destination core is in the ideal status, the physical address is sent via H-tree bus. Otherwise, the data fetching request is temporarily stalled by the arbiter and waits for the next-round allocation. 

\noindent
\textbf{Product Quantization (PQ) Module}
contains a codebook memory and total 32 FP16 multiply and accumulate (MAC) units to compute the $C\times M$ ADT. 
The codebook memory stores the $C\times D$ PQ codebook trained by offline $k$-means. Since the parameters $C=256, M=32$ are fixed for different datasets while the vector dimension $D$ ranges from 96 to 128, we use a 64kB SRAM as the codebook memory and a 16kB SRAM as the ADT memory. 
The ADT computation with $\mathcal{O}(C\cdot D)$ complexity incurs latency with $8D$ (Angular distance) to $24D$ (Euclidean distance) clock cycles.

\noindent
\textbf{Distance Computation Module} inside each queue consists of one ADT memory, one query buffer, and one MAC unit. It supports two types of distance computations: the approximate PQ distance and accurate distance. 
For the PQ distance, the distance computation module uses the fetched 256-b PQ code to look up the partial distances of $M$ sub-vectors in the ADT memory. Then it computes the accumulation (Eq. (\ref{eq:pq_dist})) in $M$ clock cycles. 
For accurate distance, the base data vectors are first fetched and computed against the query data cached in the query buffer, requiring $D$ clock cycles in total.

\noindent
\textbf{Queue} 
works independently to process the incoming data from the scheduler or the arbiter. 
The data from PQ module include the ADT and query vector, which are used to initialize the ADT memory and query buffer. 
The queue sends vertex information to the arbiter for fetching data stored in 3D NAND flash tiles. The read data (PQ codes, raw base vector, or NN indices) are sent back the corresponding queue via the arbiter. 
One key design for high-performance graph search is the search queues. \Design accelerator contains total $N_{core}=512$ cores which provide high internal data parallelism. Single queue is unable to fully leverage this parallelism. To increase the core utilization as well as search throughput, $N_q$ queues are implemented in the search queue module of Fig.~\ref{fig:search_engine}. 

\noindent
\textbf{Bloom Filter.} 
Some vertices may be visited repeatedly during graph search. To save redundant computations, the indices of visited vertices are saved and checked. Based on our simulation, the number of visited vertices increases linearly with the candidate list size $|\mathcal{L}|$. At most 8000 vertices need to be saved for $|\mathcal{L}|=250$. 
We use the memory-efficient \textit{Bloom filter}~\cite{SONG} to detect visited vertices. Bloom filter is a probabilistic data structure~\cite{bloom1970space} with a false positive of $(1-e^{kn/m})^{k}$, where $k$ is the number of hash functions, $m$ is the size bit array size, and $n$ is the number of elements to be inserted. We implement a 12kB SRAM with 8 hashes in the Bloom filter module to guarantee a false positive probability $<0.02$\%, providing negligible recall loss~\cite{SONG}.

\noindent
\textbf{Bitonic Sorter and Candidate List.} 
During the graph traversal, the candidate list in each search queue requires sorting to obtain the sorted candidate list. In the worst case, over 200 distances need to be sorted.  We use the parallel \textit{Bitonic sorter} that provides constant latency to avoid excessive sorting latency. The Bitonic sorter is stage-pipelined and each cycle accepts $N_{sorter}$ inputs in parallel. The sorting latency for $N_{sorter}$ inputs is constant $2\log_2N_{sorter}$ clock cycles. \Design only implements one global $N_{sorter}=256$-point Bitonic sorter that satisfy the latency and throughput requirements of used list size $|\mathcal{L}|$ and queues. The candidate list is a 2kB buffer to store the candidate set $\mathcal{L}$ in Algorithm~\ref{alg:ann_search}, including each candidate's distance and the corresponding vertex index. The candidate list also stores the previous $k$-NN results from the last iteration to support the early termination in Section~\ref{subsec:dynamic_et}.

\begin{figure}[t]
    \centering
    \begin{subfigure}[b]{0.9\linewidth}
        \centering
        \includegraphics[width=0.85\linewidth]{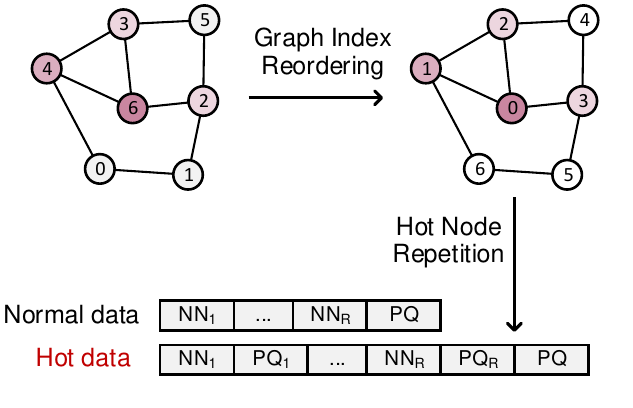}
        \caption{}
    \end{subfigure}
    \hfill
    \begin{subfigure}[b]{0.9\linewidth}
        \centering
        \includegraphics[width=0.63\linewidth]{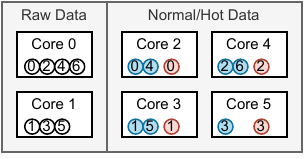}
        \caption{}
    \end{subfigure}
    \caption{{(a) Graph index reordering and hot nodes repetition. (b) Address mapping.}}
    \label{fig:data_mapping}
\end{figure}

\section{\Design Data Mapping}\label{sec:data_mapping}


\subsection{Graph Pre-processing}\label{subsec:graph_preproc}
Mapping the raw graph indices directly into 3D NAND yields sub-optimal performance due to poor locality. Prior work, such as NDSEARCH~\cite{wang2024ndsearch} improves locality through graph pre-processing, but its strategies have two key limitations. First, its degree-based static scheduling is ineffective for modern graph indices whose node degrees are nearly uniform (\eg, HNSW~\cite{hnsw}). Second, NDSEARCH ignores the highly skewed node visiting frequency observed in real workloads. 
In contrast, \Design employs hotness-based graph reordering and node repetition, which reorder and replicate frequently visited nodes to enhance die-level locality and reduce effective access latency. These optimizations are uniquely enabled by \Design's fine-grained Cu–Cu bonded NSP architecture and overcome limitations in NDSEARCH.

\noindent
\textbf{Hotness-based Graph Reordering.} 
The generated graph from ANNS tools~\cite{diskann,hnsw} has a highly random index distribution. Instead of directly mapping the irregular graph indices, we first reorder the indices to increase their locality. 
As shown in Fig.~\ref{fig:data_mapping}-(a), the graph index reordering is based on the vertices' visiting frequency. The hotter (more frequent) vertices have smaller indices, which means the entry point starts from 0. The calculation of the vertices' visiting frequency is based on the graph search trace from the randomly sampled base data. The graph index reordering is helpful to increase the index locality.

\noindent
\textbf{Hot Nodes Repetition.} 
After graph index reordering, the hot vertices and their NNs will be located on the top of graph index. 
\Design selects the hottest nodes which have the smallest indices as the hot nodes. The hot nodes repetition scheme ensures the graph search process for the most frequently accessed vertices can be significantly accelerated. 
As shown in Fig.~\ref{fig:data_mapping}-(a), each hot node's NN index is followed by the corresponding PQ code. In this case, the NNs' PQ codes and indices are stored together, so computing each vertex can be done in one shot, which means only one WL setup for 3D NAND is sufficient. 
The cost of hot nodes repetition scheme is the additional bits for $R$ PQ codes. The total bits for each hot node are $R\times(b_{index}+b_{PQ})+ b_{PQ}$.

\subsection{Gap Encoding for Vertex Indices}
The graph ANNS algorithms store the NN index and raw data. This leads to two problems: 1. Fetching vertex indices (NN indices) creates significant data access overhead during search, accounting for 80\% to 90\% as shown in Fig.~\ref{fig:convergence_traffic}-(b). 2. The NN index has a comparable size as the raw data. 
Existing algorithms~\cite{hnsw,nsg,diskann} use a uniform 32-b integer to present the vertex index. But the uniform bit width for different graph scales is redundant since $\lceil \log_2N \rceil$-b is sufficient to present the vertex index for a dataset with size $N$. 
\Design uses gap encoding~\cite{besta2018survey} to save the space and data movement required by NN indices. 
Fig.~\ref{fig:gap_pq}-(a) illustrates an example for 4 indices and 3 NNs that are expressed by an adjacency list of 12 elements. The uncompressed 32-b adjacency list consumes 384-b space. In comparison, gap encoding includes two steps: first, sort the NN indices in each row in ascending order and then convert the sorted indices into the difference values to its previous index except for the first one. In this case, the bit width is determined by the bits for the maximum difference value. Therefore, the required space is reduced to 168-b. 
Our experiments show graphs of 1M to 100M datasets need 20-b to 26-b gap encoding, leading to at least 19\% to 37\% graph index data compression. 
The compressed graph index also helps achieve faster graph traversal as the overhead of index fetching is reduced.

\subsection{Data Layout and Allocation}\label{subsec:data_layout}
Most graph search operations are related to PQ codes and graph indices, while access to raw data is needed for accurate distance computation during the reranking step. We let \Design store the raw data individually in some 3D NAND cores while PQ codes and graph indices are stored together. 
As shown in Fig.~\ref{fig:data_mapping}-(b), we use a core-level round-robin address mapping scheme, such that the data with consecutive indices are assigned to consecutive cores. This scheme helps maximize memory utilization. Each page uses the same bit length to store node vectors and associated NN indices (nodes with degree$<R$ are padded to $R$ to align address). 
Fig.~\ref{fig:data_mapping}-(a) shows the coupled storage format of PQ codes and graph indices in each page of 3D NAND core. Each data frame for one vertex $v_i$ starts with $R$ NNs' graph indices and the end is the corresponding PQ code of $v_i$. So $R\times b_{index}+b_{PQ}$ bits are needed for each vertex. Each page can store at most $\lfloor \frac{N_{BL}}{R\times b_{index}+b_{PQ}} \rfloor$ frames. Likewise, each raw data consumes $b_{raw}\times D$ bits and each page can store $\lfloor \frac{N_{BL}}{b_{raw}\times D} \rfloor$ raw data frames. 


\begin{table}[t]
    \caption{Specifications of evaluated datasets.}
    \centering
    \scriptsize
    \renewcommand{\arraystretch}{1.05}
        \begin{tabular}{c|c|c|c|c}
            \hline
            \textbf{Dataset} & \textbf{Distance} & \textbf{\# Base} & \textbf{\# Query} & \textbf{Dimension $D$} \\
            \hline
            \hline
            {SIFT}~\cite{product_quantization} & Euclidean & 1M & 10K & 128\\
            \hline
            {GLOVE}~\cite{glove} & Angular & 1M & 10K & 100\\
            \hline
            {DEEP-10M}~\cite{deep_1B} & Inner Product & 10M & 10K & 96 \\
            \hline
            {BIGANN-10M}~\cite{sift_1B} & Euclidean & 10M & 10K & 128 \\
            \hline
            {DEEP-100M}~\cite{deep_1B} & Inner Product & 100M & 10K & 96\\
            \hline
            {BIGANN-100M}~\cite{sift_1B} & Euclidean & 100M & 10K & 128 \\
            \hline
        \end{tabular}
    \label{table:dataset}
\end{table}

\section{Evaluation}
\subsection{Evaluation Methodology}
\noindent
\textbf{Baselines.} 
\Design graph search is compared with three graph-based ANNS baselines: HNSW~\cite{hnsw} and DiskANN~\cite{diskann} on CPU and GGNN~\cite{ggnn} on GPU. 
We also consider the compression-based algorithm, IVF-PQ in FAISS~\cite{faiss}, as the non-graph baseline. 
We use the graph building parameters of ANN-Benchmarks~\cite{ann_benchmark}: the maximum out degree $R=64$ with list size $L=150$ for DiskANN and $L=500$ for HNSW. 
\Design hardware is compared with three domain-specific accelerators, ANNA~\cite{anna} (IVF-PQ-based ASIC), VStore~\cite{vstore} (near-storage graph ANNS accelerator), and ICE~\cite{ice}.

\noindent
\textbf{Benchmarks.} 
Table \ref{table:dataset} lists the used ANNS datasets on three scales (1M, 10M and 100M). The 10M and 100M datasets are subsets extracted from DEEP1B~\cite{deep_1B} and BIGANN1B~\cite{sift_1B}, respectively. 
The software baselines are evaluated on an AMD EPYC 7543 CPU with 128 GB DDR4-3200 memory and NVIDIA A100 GPUs with 40~GB HBM2e memory to measure their performance, energy consumption, and recall. The datasets with 1M and 10M vectors are evaluated using a single GPU, while the 100M-scale datasets for GGNN \cite{ggnn} are evaluated using two A100 GPUs due to the large graph size.

\noindent
\textbf{\Design Software Implementation.}
We implement the \Design graph search algorithm using C++ upon DiskANN codebase~\cite{diskann}. The code is compiled with g++ compiler with \texttt{-O3} optimization. 
The data vectors are divided into 32 subvectors and 256 centroids per subvector. 
The PQ threshold is $beta=1.06$ illustrated in Section \ref{subsec:rerank}, repetition rate $r$ in the range 1 to 15 and search step $T_{step}=4$ in Section \ref{subsec:dynamic_et}.

\noindent
\textbf{\Design Hardware.}
\Design performance is estimated using an in-house simulator: The simulator's back-end is built upon 3D-FPIM~\cite{3D-FPIM} to compute the parasitics, energy, and latency of 3D NAND based on the parameters of Samsung's 96-layer NAND flash~\cite{techinsight_vnand} and configurations in Table~\ref{tab:hard_imp}. 
The simulator's front-end is a modified version of NeuroSIM~\cite{neurosim} which faithfully reflects the behavior of \Design graph search. The front-end accepts the trace generated from the software and calculates the corresponding performance metrics. 
The ASIC search engine is implemented using Verilog HDL using TSMC 40nm technology. The clock frequency is 1GHz and the design is scaled to 22nm. The timing and energy parameters of the SRAM and buffers are calculated using CACTI-3DD \cite{cacti} at 22nm.

\begin{table}[t]
    \caption{Area and power breakdown of \Design accelerator.}
    \centering
    \scriptsize
    \renewcommand{\arraystretch}{1.02}
    \resizebox{\linewidth}{!}{
    \begin{tabular}{c|c|c|c|c|c}
        \hline
        & \multirow{2}{*}{\textbf{Hardware Unit}} & \multicolumn{2}{c|}{\multirow{2}{*}{\textbf{Configs.}}}  & \textbf{Area}  & \textbf{Dynamic} \\
        & & \multicolumn{2}{c|}{} & (mm$^2$) & \textbf{Energy} (pJ) \\
        \hline
        \hline
        \multirow{7}{*}{\rotatebox[origin=c]{90}{\textbf{3D NAND Flash}}} & \textbf{Core} & \multicolumn{2}{c|}{$\times 1$} & 0.505 & - \\
        \cline{2-6}
        &  3D NAND Blocks & \multicolumn{2}{c|}{96-layer, 4 SSL, 36864 BL}  & 0.505 & 4442. \\
    & Core H-Tree Bus & \multicolumn{2}{c|}{$\times 1$} & 0.163 & 21.4 \\        
        \cline{2-6}
        & \textbf{Tile} & \multicolumn{2}{c|}{$\times 1$} & 16.16 & - \\
        \cline{2-6}
        & Core & \multicolumn{2}{c|}{$\times 32$} & 16.16 & - \\
        & Tile H-Tree Bus & \multicolumn{2}{c|}{$\times 1$} & 1.309 & 198.6 \\
        \cline{2-6}
        & \textbf{Total} & \multicolumn{2}{c|}{16 Tiles (432Gb)} & 258.56 & - \\
        \hline
        \hline
        & \multirow{2}{*}{\textbf{Hardware Unit}} & \multirow{2}{*}{\textbf{Size}}  & \textbf{Area}  & \textbf{Dynamic} & \textbf{Static} \\
        & & & (mm$^2$) & \textbf{Pwr.} (mW) & \textbf{Pwr.} (mW)\\
        \hline
        \multirow{9}{*}{\rotatebox[origin=c]{90}{\textbf{Search Engine}}} & \textbf{Search Queues}  & $\times 256$ & 9.012 & 1920.316 & 2127.384 \\
        \cline{2-6}
        & Candidate List & 2kB$\times 1$ & 0.003 & 0.274 & 0.684  \\
        & Bloom Filter & 12kB$\times 1$ & 0.014 & 4.579 & 3.472  \\
        & ADT Module & 16kB$\times 1$ & 0.017 & 1.793 & 4.153  \\
        \cline{2-6}
        & \textbf{PQ Module} & $\times 1$ & 0.082 & 17.396 & 14.347 \\
        \cline{2-6}
        & Codebook Mem. & 64kB$\times 1$ & 0.058 & 5.822 & 14.345 \\
        & FP16-MACs & $\times 32$ & 0.024 & 11.574 & 0.002   \\
        \cline{2-6}
        & \textbf{Bitonic Sorter} & $\times 1$ & 0.237 & 486.090 & 0.021 \\
        \cline{2-6}
        & \textbf{Total} & - & 9.331 & 2423.802 & 2141.752 \\
        \hline
    \end{tabular}
    }
    \label{tab:hard_imp}
\end{table}

\subsection{ANNS Algorithm Evaluation}

\begin{figure}[t]
    \centering
    \includegraphics[width=\linewidth]{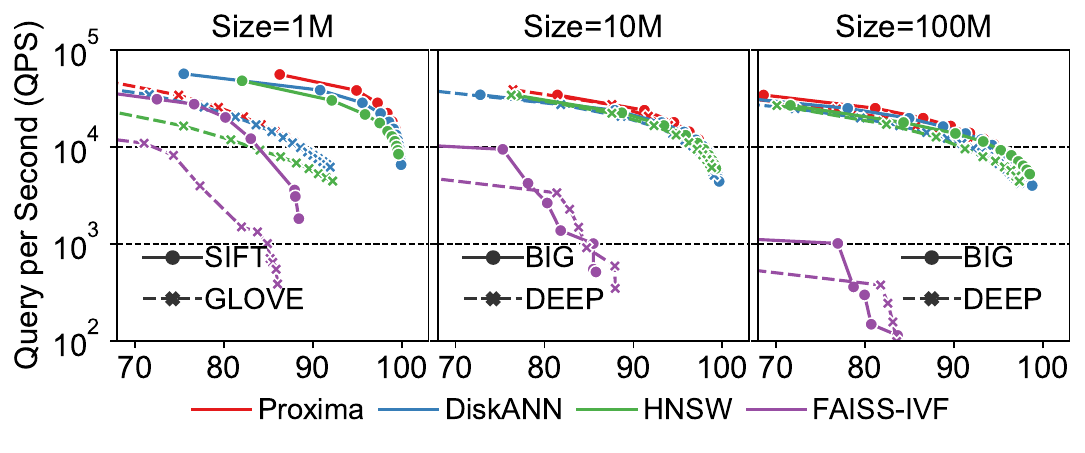}
    \caption{Throughput (QPS) and recall comparison with HNSW~\cite{hnsw}, DiskANN~\cite{diskann}, and FAISS-IVF~\cite{faiss} on CPU.}
    \label{fig:software_recall_qps}
\end{figure}

\noindent
\textbf{Improvements over Graph Baselines.}
We evaluate \Design graph search and compare it with DiskANN~\cite{diskann} and HNSW~\cite{hnsw} on CPU. Fig.~\ref{fig:software_recall_qps} shows throughput (QPS) and recall, where \Design achieves comparable recall and throughput across all datasets and up to 10\% recall improvement over DiskANN and HNSW at the same throughput on 1M datasets. The improvements over HNSW come from using lightweight PQ distance approximations instead of expensive accurate distances. Compared to DiskANN that also uses PQ, such improvements come from our implementation of the novel search strategy in Algorithm~\ref{alg:ann_search}.

\noindent
\textbf{Comparison with Non-graph Baseline.} 
We also evaluate the non-graph algorithm, FAISS-IVF~\cite{faiss}, which uses PQ compression. Although the memory footprint of FAISS-IVF is smaller, lossy PQ compression introduces significant approximation error and causes the recall to saturate around 90\% and 85\% for small and large datasets, respectively. Graph-based methods consistently achieve better recall vs. throughput tradeoff over FAISS-IVF.

\begin{figure}[t]
    \centering
    \begin{subfigure}[b]{\linewidth}
         \centering
         \includegraphics[width=\linewidth]{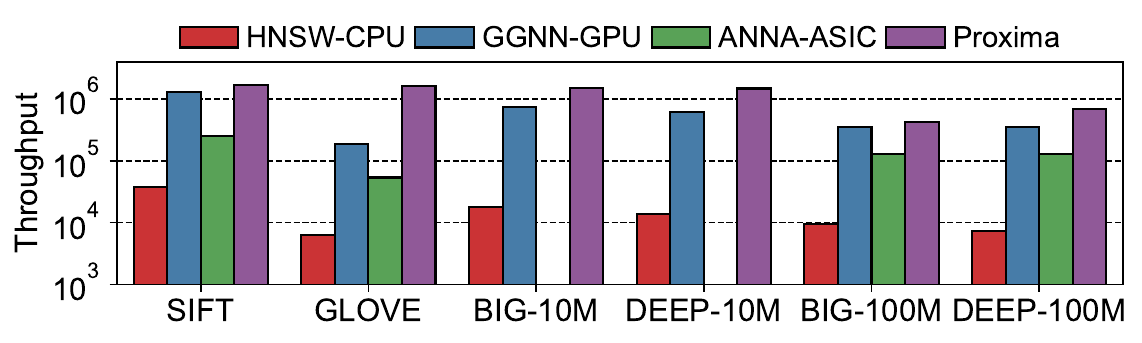}
         \caption{Throughput (QPS)}
     \end{subfigure}
     \hfill
     \begin{subfigure}[b]{\linewidth}
         \centering
         \includegraphics[width=\linewidth]{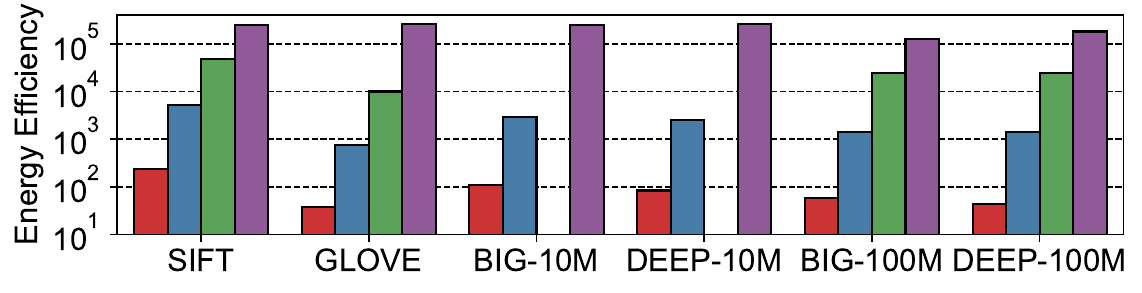}
         \caption{Energy efficiency (QPS/W)}
     \end{subfigure}
    \caption{Throughput and energy efficiency comparison for HNSW~\cite{hnsw}, GGNN~\cite{ggnn}, ANNA~\cite{anna}, and \Design.}
    \label{fig:comp_hardware_qps_energy}
\end{figure}

\begin{figure*}[t]
    \centering
    \includegraphics[width=\linewidth]{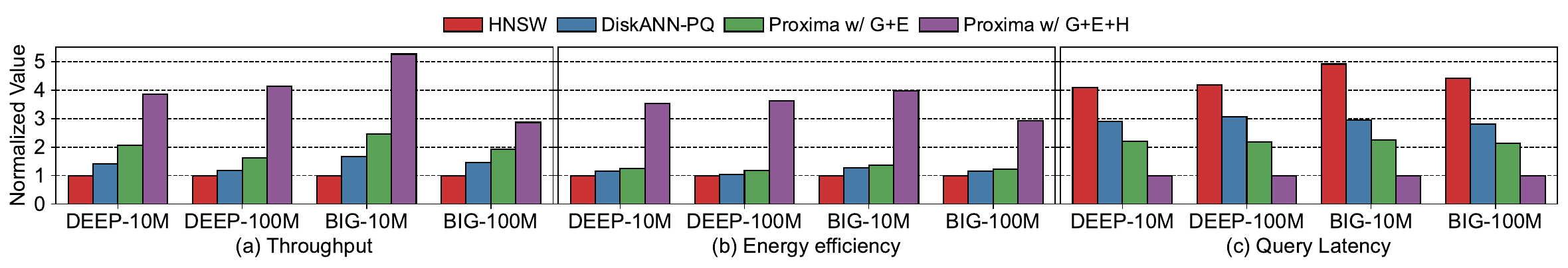}
    \caption{Performance comparison for various graph ANNS algorithms ran on the proposed 3D NAND-based NSP accelerator: HNSW~\cite{hnsw}, DiskANN-PQ~\cite{diskann}, and \Design (G: gap encoding, E: early termination, H: hot nodes repetition).}
    \label{fig:graph_perf_hardware}
\end{figure*}

\subsection{Hardware Performance Comparison}

\noindent
\textbf{\Design vs. CPU, GPU, and ASIC.}
We compare \Design near-storage accelerator in 3D NAND with state-of-the-art CPU, GPU, and ASIC designs for ANNS. 
Fig.~\ref{fig:comp_hardware_qps_energy}-(a) summarizes the throughput comparison, which is conducted on two small-scale datasets (SIFT and GLOVE) and two large-scale 100M datasets (BIGANN-100M and DEEP-100M). 
GGNN~\cite{ggnn} is a high-performance GPU tool optimized for graph-based ANNS while ANNA~\cite{anna} is a PQ-IVF-based ASIC accelerator. We adopt HNSW~\cite{hnsw} as the CPU baseline. 
Proxima achieves the highest throughput among all baselines while GPU-based GGNN is the 2nd fastest tool. Compared to ASIC design, ANNA~\cite{anna}, Proxima is $6.6\times$ to $13\times$ faster on 1M and 100M datasets. 
Compared to the CPU baseline, \Design's speedup is more significant for large-scale datasets or high-complexity datasets, such as GLOVE~\cite{glove} that needs much more distance computations ($6\times$ to $8\times$) to achieve the same recall.

We measure the energy efficiency for \Design and the baselines mentioned in Fig.~\ref{fig:comp_hardware_qps_energy}-(b). Energy efficiency is measured by QPS/W. \Design has the highest energy efficiency in all datasets. ANNA's energy efficiency is up to $17\times$ inferior to \Design because ANNA needs large on-chip SRAMs and frequent off-chip data transfer. The expensive data movement increases the energy consumption. Compared to CPU and GPU accelerators, \Design is two to three orders of magnitude energy efficient because \Design implemented near 3D NAND has much shorter data access latency and more energy-efficient datapath due to the adopted NSP technique.

\noindent
\textbf{Hardware Overhead.} 
Table~\ref{tab:hard_imp} summarizes the area and energy breakdown for \Design near-storage accelerator. The area of the 3D NAND part is dominated by the memory tier determined by the size of the memory array. The areas of the peripheral circuits and H-Tree buses are factored out by incorporating the heterogeneous integration. The search engine area is also factored out by fitting the search engine on the top CMOS tier, The overall \Design area is 258.56 mm$^2$, which is a compact solution for graph-based ANN. \Design' area size is $2.4\times$ smaller than the A40 GPU's 628mm$^2$ die area.

\begin{table*}[ht]
    \caption{Comparison with existing CPU, GPU, ASIC, and NSP accelerators.}
    \centering
    \scriptsize
    \renewcommand{\arraystretch}{1.25}
        \begin{tabular}{c|c|c|c|c|c}
            \hline
            \textbf{Design} & DiskANN-PQ~\cite{diskann} & GGNN~\cite{ggnn} & ANNA~\cite{anna} &  VStore~\cite{vstore} & \Design \\
            \hline
            \hline
            \textbf{Platform} & CPU & GPU & ASIC & FPGA+SSD  & 3D NAND SLC\\
            \hline
            \textbf{Including storage?} & No & No & No & \textbf{Yes} & \textbf{Yes}\\
            \hline
            \textbf{Memory Type} & DRAM-DDR4-3200 & HBM2e & DRAM & DRAM+SSD & 3D NAND\\
            \hline
            \textbf{Memory Capacity} & 128GB & 32GB & - & 32GB & 54GB\\
            \hline
            \textbf{Peak Bandwidth} & 102GB/s & 900GB/s& 64GB/s & 9.9GB/s (aggregated) & 254GB/s \\
            \hline
            \textbf{Memory Bit Density} & 0.2Gb/mm$^2$ & 0.7Gb/mm$^2$ & 0.2Gb/mm$^2$ & 4.2Gb/mm$^2$ & 1.7Gb/mm$^2$ \\
            \hline
        \end{tabular}
    \label{table:hardware_comparison}
\end{table*}

\subsection{Effectiveness of \Design Optimizations}

\begin{figure}[t]
    \centering
    \includegraphics[width=0.95\linewidth]{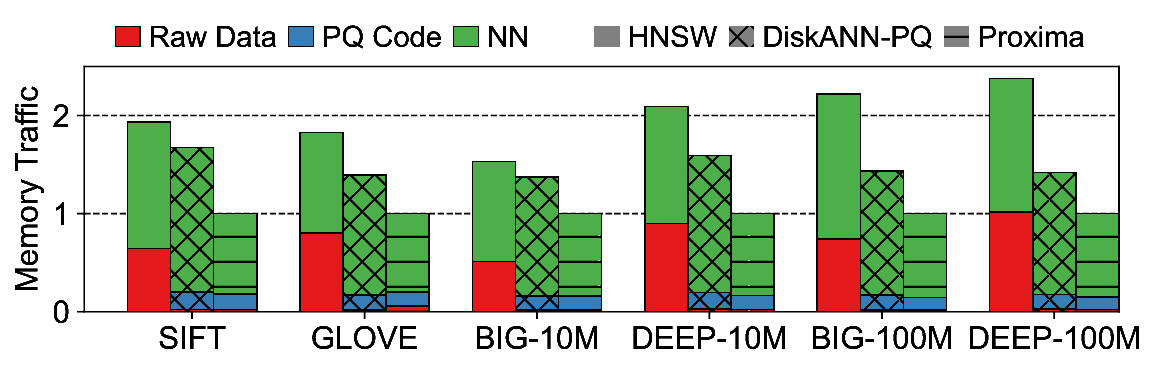}
    \caption{Memory traffic breakdown for HNSW~\cite{hnsw}, DiskANN-PQ~\cite{diskann}, and \Design with gap encoding and early termination.}
    \label{fig:data_traffic}
\end{figure}

\noindent
\textbf{Gap Encoding and Early Termination.} We analyze the memory traffic for different graph ANNS search algorithms in Fig.~\ref{fig:data_traffic}. HNSW~\cite{hnsw} without using any distance approximations incurs the largest data movement overhead. DiskANN-PQ denotes DiskANN with product quantization~\cite{diskann}. DiskANN-PQ reduces 12\% to 40\% total memory traffic because most raw data access is skipped using PQ. 
Gap encoding in \Design further decreases the data access of NN indices, while early termination saves 10\% to 25\% redundant operations over DiskANN-PQ. As a result, \Design achieves a reduction ratio of $1.9\times$ to $2.4\times$ over HNSW~\cite{hnsw}.

\begin{wrapfigure}{r}{0.6\linewidth}
    \centering
    \includegraphics[width=\linewidth]{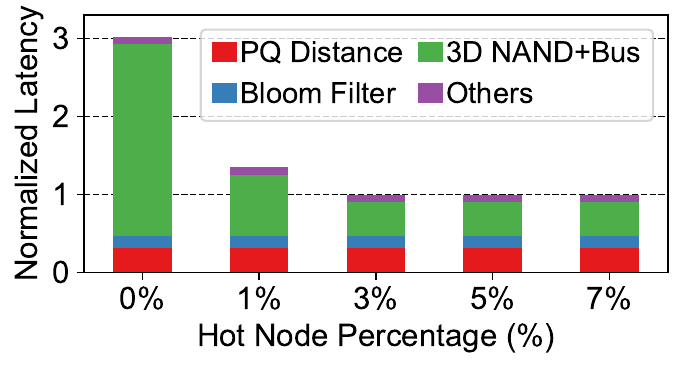}
    \caption{Runtime breakdown for different hot node percentages.}
    \label{fig:study_hot_nodes}
\end{wrapfigure}

\noindent
\textbf{Hot Nodes Repetition.} 
The hot node repetition is an effective scheme to reduce the latency while improving the overall efficiency. 
Fig.~\ref{fig:study_hot_nodes} illustrates the runtime breakdown for hot node percentages from 0.0\% to 7.0\% on 100M datasets. When hot node repetition is disabled, data access latency caused by the 3D NAND core and H-tree bus contributes to $80\%$ of the overall latency. Adding 1\% of hot nodes (1M points for 100M datasets) helps to reduces the overall search latency by 2.2$\times$. The gain comes from: the increased data locality of hot nodes helps to reduce the required data access time to the 3D NAND cores as the one round of NN indices fetching and PQ distance computations (Line 6-9 in Algorithm~\ref{alg:ann_search}) can be finished within one page access. 
When the percentage of hot nodes further increases to 3\%, the speedup is $\approx3\times$ compared to the performance without hot nodes. However, the benefits of adding hot nodes $>3\%$ reach a plateau. We use $3\%$ as the default value in \Design.

\noindent
\textbf{Improvements over other Graph ANNS Algorithms.}
\Design accelerator is general to support various graph ANNS algorithms. 
We compare the performance of two variants of \Design with two state-of-the-art works, HNSW~\cite{hnsw} and DiskANN-PQ~\cite{diskann} . All these graph algorithms run on the proposed 3D NAND accelerator using the same hardware configurations in Table~\ref{tab:hard_imp}. 
Fig.~\ref{fig:graph_perf_hardware} shows the throughput, energy efficiency, query latency on 10M and 100M datasets.  
\Design with gap encoding and early termination achieves moderate throughput and efficiency improvements over HNSW and DiskANN-PQ via software optimization. The early termination and gap encoding help to reduce both computational redundancy and data movement during search. HNSW yields the worst performance because it requires many accurate distance computations. 
After adopting the hot nodes repetition, \Design delivers about 2$\times$ speedup, 3$\times$ energy efficiency improvement, and 3$\times$ latency reduction. The benefits mainly result from the better data locality due to graph index reordering and hot nodes repetition, which reduce the memory access during graph traversal.

\subsection{Scalability and Sensitivity Analysis}\label{subsec:analysis_error}

\begin{figure}[t]
    \centering
    \includegraphics[width=0.95\linewidth]{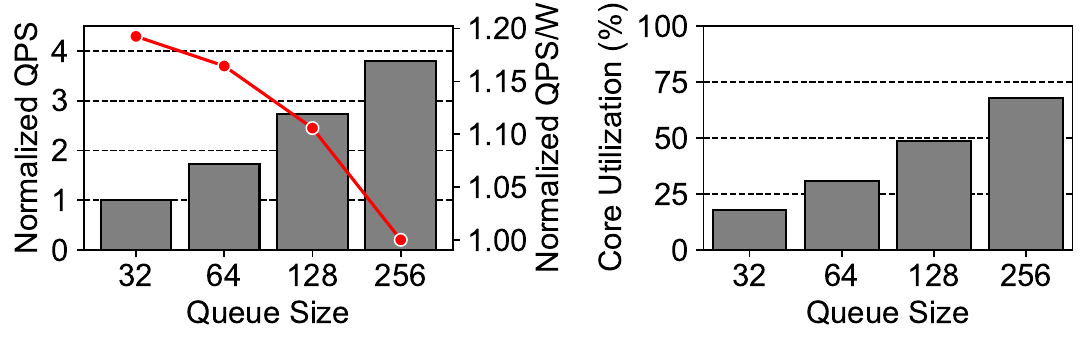}
    \caption{Performance and efficiency of \Design using different queue sizes on 100M datasets.}
    \label{fig:study_queue_size}
\end{figure}

\noindent
\textbf{Queue Size.} 
The search queue design is the key to improving throughput and the utilization of the core and bandwidth. We vary the queue size $N_q$ from 32 to 256 and simulate the normalized throughput, energy efficiency, and 3D NAND core utilization on 100-M datasets without using hot node repetition. Fig.~\ref{fig:study_queue_size} shows that increasing the queue size to 256 significantly improves the throughput by 3.8$\times$. This is because more parallel search queues can handle more queries simultaneously, thereby increasing the achievable number of parallel memory requests. Increased memory requests increase core utilization from 17.9\% to 68\%. However, increasing queue size also decreases energy efficiency by almost 20\% due to two factors: 1. The increased queues may conflict when two different queues access the same core, which slows down the query search. 2. More search queues dissipate more static power during graph search. We observe that further increasing the queue size beyond $N_q=256$ does not significantly speed up QPS while adding more power consumption to the system. This suggests that the queue has basically saturated the internal bandwidth of 3D NAND core. Therefore, it is not cost-effective to further increase the queue size. We choose $N_q=256$ as the default queue size.

\begin{figure}[t]
    \centering
    \includegraphics[width=0.95\linewidth]{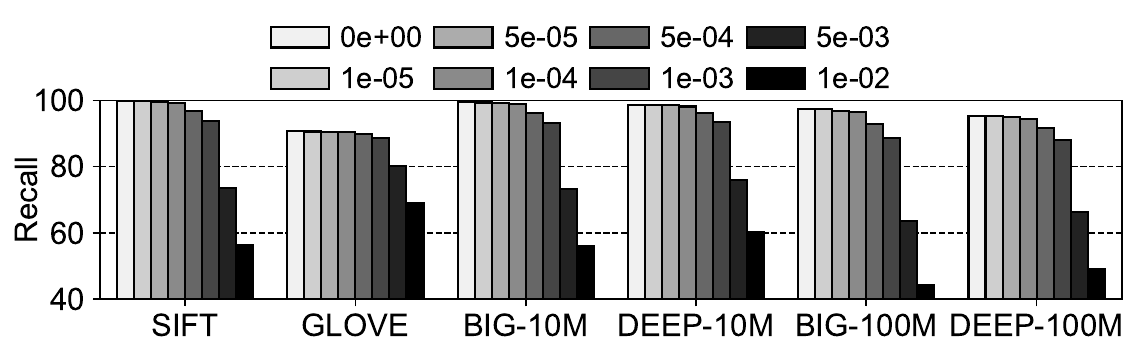}
    \caption{Impact of data errors in 3D NAND flash on the search recall across different datasets.}
    \label{fig:ber_vs_recall}
\end{figure}

\noindent
\textbf{3D NAND Error.} \Design uses SLC-based 3D NAND without any ECC modules to deliver the best. It is critical to evaluate and study whether the ECC-free design is able to tolerate the occurred errors. Typically, the raw bit error rate of SLC 3D NAND is lower than $1\mathrm{e}^{-5}$~\cite{RBER_SLC}, while larger than $1\mathrm{e}^{-4}$ for MLC 3D NAND~\cite{RBER_MLC} and TLC 3D NAND~\cite{RBER_TLC}. Fig.~\ref{fig:ber_vs_recall} shows the simulation results of search recall degradation with the impact of bit errors. It shows that using SLC 3D NAND, \Design can retain the recall rate without ECC because the recall degradation is less than $3\%$ when the error rate increases to $1\mathrm{e}^{-4}$.

\subsection{Comparison with Other ANNS Accelerators}
We first compare the key design specifications of \Design to state-of-the-art ANNS accelerations in Table~\ref{table:hardware_comparison}, including CPU-based DiskANN-PQ~\cite{diskann}, GPU-based GGNN~\cite{ggnn}, ASIC-based ANNA~\cite{anna}, and NSP-based VStore~\cite{vstore}. 
\Design and VStore are the only two designs that can persist the graph data in 3D NAND flashes. No off-chip communication is needed for the next-time power on. 
In comparison, GGNN and ANNA rely on off-chip or on-chip data communication, which incurs higher energy consumption. 
VStore relied on SSD internal interface like SimpleSSD to realize the data communication with processors, which delivers very low bandwidth (9.9GB/s). The low bandwidth severely limits scaling to large-scale graph ANNS applications. 
Compared to VStore, \Design has more compact design, shorter memory latency, and 26$\times$ higher peak memory bandwidth due to the high-speed heterogeneous integration. 
GGNN with NVIDIA V100 GPU and HBM2e memory delivers the highest bandwidth. But its bit density is only 41\% of \Design.
Overall, \Design balances well between memory capacity, density, and bandwidth, which can be regarded as a promising solution for data-intensive graph ANNS workloads.

\begin{table}[ht]
    \centering
    \scriptsize
    \caption{Billion-scale ANNS performance comparison with other NMP and NSP accelerators.}
    \label{table:sota_nmp_nsp_comp}
    \renewcommand{\arraystretch}{1.1}
    \setlength{\tabcolsep}{3.5pt}
    \resizebox{\linewidth}{!}{
        \begin{tabular}{c|c|c|c|c|c}
            \hline
            \textbf{Design} & Pyramid~\cite{zhu2023processing} & NDSEARCH~\cite{wang2024ndsearch} & REIS~\cite{chen2025reis} & VStore~\cite{vstore} & \Design $8\times$ \\
            \hline
            \hline
            \textbf{Mem. Type} & DRAM & \multicolumn{3}{|c|}{Smart SSD} & 3D NAND \\
            \hline
            \textbf{Algorithm} & \multicolumn{2}{|c|}{Graph} & IVF & \multicolumn{2}{|c}{Graph} \\
            \hline
            \textbf{Technique} & NMP & \multicolumn{3}{|c|}{NSP} & NSP + HB\\
            \hline
            \textbf{Dataset} & \multicolumn{3}{|c|}{BIGANN-1B} & DEEP-1B & BIGANN-1B \\
            \hline
            \textbf{Throughput } & 44.2k & 4.2k & 4.6k & 4.7k & 84.3k \\
            (QPS) & (10.5$\times$) & (1.0$\times$) & (1.1$\times$) & (1.1$\times$) & (20.1$\times$) \\
            \hline
        \end{tabular}
    }
\end{table}

To study the ANNS performance of \Design on billion-scale benchmarks, we scale \Design by $8\times$ to provide enough capacity for billion-scale datasets. Table~\ref{table:sota_nmp_nsp_comp} compares \Design's projected billion-scale ANNS performance with state-of-the-art NMP and NSP designs on DRAM and SmartSSD platforms~\cite{zhu2023processing, wang2024ndsearch, chen2025reis, vstore}. DRAM-based NMP (\eg, Pyramid~\cite{zhu2023processing}) offers higher throughput than other Smart SSD-based NSP designs, but suffers from expensive storage cost for billion-point graphs. SmartSSD NSP solutions (\eg, NDSEARCH~\cite{wang2024ndsearch}, REIS~\cite{chen2025reis}, and VStore~\cite{vstore}) improve density yet remain constrained by channel-level or chip-level bandwidth, yielding only 4-5k QPS. In contrast, \Design on 3D NAND with hybrid bonding achieves the highest 84.3k QPS, delivering up to 20$\times$ higher throughput than prior NSP solutions. This gain comes from \Design's fine-grained, die-level access enabled by Cu-Cu bonding, which removes SSD channel bottlenecks and unlocks the internal parallelism required for billion-scale graph traversal.

\section{Related Works}
\subsection{Large-scale ANNS}
Various approaches are proposed to address large-scale ANNS. Compression-based algorithms compress data based on space partitioning and clustering including IVF~\cite{ivf}, product quantization \cite{product_quantization} and hash methods \cite{charikar2002similarity}. The previous experiments~\cite{ann_benchmark} show that the low memory comsumption of these compressioncomesed methods come at the cost of moderate accuracy degradation especially on large-scale datasets. 
Graph-based methods~\cite{diskann,nsg} build an approximate graph and prune edges by proximity graphs, showing promising query speed, accuracy, and scaling performance on large datasets. These works can benefit from \Design to enhance their performance because most existing graph methods share the same graph search strategy.

\subsection{Acceleration for Graph-based ANNS}
There exist various hardware accelerators that aim at addressing large-scale ANNS problems with high energy efficiency as well as low query latency. These works include VStore~\cite{vstore}, ICE~\cite{ice}, SSAM~\cite{lee2018application}, and CXL-ANNS~\cite{jang2023cxl}. 
Near-storage processing~\cite{vstore,lee2018application,jang2023cxl,wang2024ndsearch,chen2025reis} or in-storage processing~\cite{jang2023cxl,ice} is commonly utilized to achieve better system efficiency and high storage density on vairous types of memory devices. 
\Design co-designs the graph search algorithm and the 3D NAND-based hardware, thereby having better memory density, energy efficiency, and bandwidth trade-offs as compared to existing works.

\section{Conclusion}
This paper presents \Design, an NSP-based accelerator for graph-based ANNS. Based on an analysis of existing ANNS tools, we characterize that the algorithm faces challenges in memory consumption, expensive distance computation, and irregular data access. 
Our solution significantly reduces the computational complexity with approximation and early termination of distance computation while showing comparable recall to competing tools. Furthermore, we devise a 3D NAND flash-based NSP accelerator that efficiently processes the optimized ANNS algorithm with strategies to maximize internal parallelism and bandwidth utilization. Compared to these state-of-the-art ANNS accelerator, \Design achieves one to two orders of magnitudes speedup and energy efficiency improvements.

\section*{Acknowledgments}
This work was supported in part by PRISM and CoCoSys, centers in JUMP 2.0, an SRC program sponsored by DARPA; SRC Global Research Collaboration (GRC) grant; and National Science Foundation (NSF) grants \#1826967, \#1911095, \#2052809, \#2112665, \#2112167, and \#2100237.

\ifCLASSOPTIONcaptionsoff
  \newpage
\fi

{
    \tiny
\bibliographystyle{IEEEtran}
    \bibliography{refs}
}

\vskip -2.5\baselineskip plus -1fil
\begin{IEEEbiographynophoto}{Weihong Xu}
received his B.E. and M.E. degrees from Southeast University in 2017 and 2020, respectively, and his Ph.D. degree from the University of California, San Diego (UCSD) in 2024. He was a postdoctoral researcher at the École Polytechnique Fédérale de Lausanne (EPFL) in 2025.
He is currently a ZJU100 Young Professor in the College of Integrated Circuits at Zhejiang University. His research interests include heterogeneous near-memory computing, RISC-V–based processor and accelerator architectures, and algorithm–hardware co-design for AI acceleration, large language models (LLMs), retrieval-augmented generation (RAG), and other emerging data-intensive workloads.
\end{IEEEbiographynophoto}

\vskip -2.5\baselineskip plus -1fil
\begin{IEEEbiographynophoto}{Junwei Chen}
is a fourth year undergraduate student in Computer Science and Mathematics at University of California San Diego, graduating in 2024, and an incoming master's student at Carnegie Mellon University. His research interests include algorithm-hardware co-design and domain-specific accelerators.
\end{IEEEbiographynophoto}

\vskip -2.5\baselineskip plus -1fil
\begin{IEEEbiographynophoto}{Po-kai Hsu}
received the B.S. degree in electrical engineering
and the M.S. degree in electro-optical engineering from National Tsing Hua University, Hsinchu,
Taiwan, in 2014 and 2018, respectively. He is currently pursuing the Ph.D. degree in electrical and
computer engineering with the Georgia Institute of
Technology, Atlanta, GA, USA. From 2018 to 2021,
he was a Device Engineer with the Emerging Central Laboratory, Macronix International Company
Ltd., Hsinchu. His current research interests include
algorithm-hardware co-design for the hardware accelerators of genome
sequencing applications and large-language models.
\end{IEEEbiographynophoto}

\vskip -2.5\baselineskip plus -1fil
\begin{IEEEbiographynophoto}{Jaeyoung Kang}
is a Ph.D. Candidate in Electrical and Computer Engineering at the University of California San Diego, La Jolla, CA, USA. He received a B.E. degree in electrical engineering from Korea University, Seoul, South Korea, in 2019. His research interests include deep learning-based algorithm acceleration on heterogeneous system architecture, GPU-based acceleration for big data analysis in bioinformatics, and hyperdimensional computing.
\end{IEEEbiographynophoto}

\vskip -2.5\baselineskip plus -1fil
\begin{IEEEbiographynophoto}{Minxuan Zhou}
is currently a Ph.D. candidate in the Department of Computer Science and Engineering at the University of California San Diego. He received his bachelor's degree from Beihang University in 2015 and his master's degree from the University of California San Diego in 2017. His research interests include computer architecture, software-hardware co-design, and domain-specific acceleration.
\end{IEEEbiographynophoto}

\vskip -2.5\baselineskip plus -1fil
\begin{IEEEbiographynophoto}{Sumukh Pinge}
is a 2nd-year Ph.D. student in Computer Science and Engineering at the University of California San Diego, La Jolla, CA, USA, affiliated with the Systems Energy Efficiency Lab (SEELab). He received his B.E. in Electronics and Instrumentation from the Birla Institute of Technology and Science (BITS), Pilani, in 2022. His research focuses on near-storage and FPGA-based hardware accelerators for bioinformatics, mass spectrometry applications, and hyperdimensional computing-based workloads.
\end{IEEEbiographynophoto}

\vskip -2.5\baselineskip plus -1fil
\begin{IEEEbiographynophoto}{Shimeng Yu}
(Fellow, IEEE) received the B.S. degree in microelectronics from Peking University in 2009 and the M.S. and Ph.D. degrees in electrical engineering from Stanford University in 2011 and 2013, respectively. From 2013 to 2018, he was an Assistant Professor in electrical and computer engineering with Arizona State University. He is currently a Full Professor in electrical and computer engineering with the Georgia Institute of Technology. His research interests include nanoelectronic devices and circuits for energy-efficient computing systems. His expertise is on the emerging non-volatile memories (e.g., RRAM and ferroelectrics) for different applications, such as machine/deep learning accelerator, neuromorphic computing, monolithic 3-D integration, and hardware security. He was a recipient of the NSF Faculty Early Career Award in 2016, the IEEE Electron Devices Society (EDS) Early Career Award in 2017, the ACM Special Interests Group on Design Automation (SIGDA) Outstanding New Faculty Award in 2018, the Semiconductor Research Corporation (SRC) Young Faculty Award in 2019, the ACM/IEEE Design Automation Conference (DAC) Under-40 Innovators Award in 2020, the IEEE Circuits and Systems Society (CASS) Distinguished Lecturer from 2021 to 2022, and the IEEE EDS Distinguished Lecturer from 2022 to 2023.
\end{IEEEbiographynophoto}

\vskip -2.5\baselineskip plus -1fil
\begin{IEEEbiographynophoto}{Tajana Rosing}
received her Ph.D. degree from Stanford University, Stanford, CA, USA, in 2001. She is a Professor, a Holder of the Fratamico Endowed Chair, and the Director of System Energy Efficiency Laboratory, University of California at San Diego, La Jolla, CA, USA. From 1998 to 2005, she was a full-time Research Scientist with HP Labs, Palo Alto, CA, USA, while also leading research efforts with Stanford University, Stanford, CA, USA. She was a Senior Design Engineer with Altera Corporation, San Jose, CA, USA. She is leading a number of projects, including efforts funded by DARPA/SRC JUMP 2.0 PRISM program with focus on design of accelerators for analysis of big data, DARPA and NSF funded projects on hyperdimensional computing, and SRC funded project on IoT system reliability and maintainability. Her current research interests include energy-efficient computing, cyber–physical, and distributed systems.
\end{IEEEbiographynophoto}




\end{document}